%% file: Layout.tex
\documentclass[format=acmsmall, review=false, screen=true]{acmart}

\usepackage[T1]{fontenc}
\usepackage{currvita}
\usepackage{booktabs} 
\usepackage{multirow}
\usepackage{subcaption}
\usepackage[ruled]{algorithm2e} 

\SetAlFnt{\small}
\SetAlCapFnt{\small}
\SetAlCapNameFnt{\small}
\SetAlCapHSkip{0pt}
\IncMargin{-\parindent}

\usepackage{array}

\makeatletter

\providecommand{\tabularnewline}{\\}

\makeatother

\usepackage[english]{babel}
\acmJournal{CSUR}
\acmVolume{0}
\acmNumber{0}
\acmArticle{00}
\acmYear{0000}
\acmMonth{00}
\copyrightyear{0000}

\setcopyright{acmlicensed}

\acmDOI{xxx}

\received{Apr. 2019}
\received[Jul.]{2019}

\begin{document}
\title[Blockchain Literature Review]{Blockchains: a Systematic Multivocal Literature Review}

\author{Bert-Jan Butijn}
\orcid{0000-0002-8848-3494}
\affiliation{%
  \institution{Tilburg University}
  \streetaddress{Warandelaan 2}
  \city{Tilburg}
  \postcode{5037 AB}
  \country{The Netherlands}}
\email{b.j.butijn@uvt.nl}
\author{Damian A. Tamburri}
\affiliation{%
  \institution{Eindhoven University of Technology}
  \streetaddress{De Rondom 70}
  \postcode{5612 AP}
  \city{Eindhoven}
  \country{The Netherlands}
}
\email{d.a.tamburri@uvt.nl}
\author{Willem-Jan van den Heuvel}
\affiliation{%
 \institution{Tilburg University}
 \streetaddress{Warandelaan 2}
 \postcode{5037 AB}
 \city{Tilburg}
 \country{The Netherlands}}
\email{W.J.A.M.v.d.Heuvel@jads.nl}

\begin{abstract}

Blockchain technology has gained tremendous popularity both in practice and academia. The goal of this article is to develop a coherent overview of the state of the art in blockchain technology, using a \emph{systematic} (i.e., protocol-based, replicable), \emph{multivocal} (i.e., featuring both white and grey literature alike) literature review, to (1) define blockchain technology (2) elaborate on its architecture options and (3) trade-offs, as well as understanding (4) the current applications and challenges, as evident from the state of the art. We derive a systematic definition of blockchain technology, based on a formal concept analysis. Further on, we flesh out an overview of blockchain technology elaborated by means of Grounded-Theory.
\end{abstract}


%
%

%
%


\maketitle

\renewcommand{\shortauthors}{Butijn et al.}

\input{bodyart}

\end{document}

%% file: bodyart.tex
\section{Introduction}
In 2008 Satoshi Nakamoto, a person or group of people\footnote{To this day the identity or identities of Satoshi Nakamoto remains unkown \cite{yaga2018blockchainoverview}.}, introduced the concept of a peer-to-peer (P2P) version of electronic cash that allows for online payments to be made directly from one party to another without any trusted financial institution \citep{nakamoto2008bitcoin}. The Nakamoto article preludes the rise of Bitcoin, ushering in the dawn of blockchain technology (BCT). BCT utilizes the concept of a digital distributed ledger (i.e., a record or ``book" of transactions), which enables the participants of a P2P network to record transactions that are publicly verifiable. In essence, BCT ensures trust between parties without any trusted intermediary when performing transactions \cite{weber2016untrusted}. BCT has gained considerable scholarly attention \cite{yli2016current} --- for this very reason, we seize the opportunity to conduct and report a systematic and \emph{multivocal} study of the state of the art in BCT for the benefit of further well-founded research as well as practice. As said, we operate a multivocal systematic study, namely, we not only focus on research literature but consider the so-called \emph{grey} literature (i.e., books, technical reports, whitepapers, and more) which may carry important information concerning the BCT software architecture landscape \cite{perroud}.

We flesh out the results of our study starting from a rigorous definition of what BCT is and is not. Indeed, the terms BCT and distributed ledger technology (DLT) are frequently used interchangeably despite attempts to semantically discern them on their distinctive underlying architectures \cite{globalblockchain2017}. Second, using the well-known 4+1 software architecture framework by Kruchten \cite{kruchten19954+} as a lens for analysis, the study outlines the design options available in the state of the art for BCT architectures.

Third, through the scenario perspective several applications are presented to illustrate how BCT could be harnessed for different scenarios. By categorizing and summarizing the current applications of BCT practitioners gain more insight in the rich palette of possibilities BCT has to offer. Fourth, with this study we highlight the properties of BCT and elucidate their arising trade-offs. Insights on these properties an their trade-offs aids practitioners in making design choices while providing scholars means to assess blockchain architecture.

Fifth, the research delineates a comprehensive and data-driven overview of the challenges in the field of BCT. Following, this study thoroughly discusses the relation amongst BCT concepts based on a rigorous and systematic analysis of the literature. Establishing this relations can help practitioners further develop BCT while scholars are provided with more accurate measurements to gauge its benefits. The discussion is further strengthened by presenting highlights and observations in-depth attained from the papers under review. Finally, this research presents a systematic overview of all current research gaps to help direct future research endeavors.

\subsection{Related Work}
Several previous related surveys exist in the state of the art, even though none of them have the scope, breadth, and width we adopt in our research design. We report the most closely related here below and highlight the novelty of our work. Yli-Huumo et al. \cite{yli2016current} review and map the extant literature to indicate research gaps. This review however, predominantly features literature related to Bitcoin and corresponding issues. On one hand, the synthesis operated by Yli-Huumo et al. is only loosely systematic and, on the other hand, the field of BCT has been rapidly developing since the publication of their study \cite{yli2016current}. By comparison, our work also fleshes out the architecture of other blockchain networks, and provides a more up-to-date and systematic overview of BCT developments.

In \cite{tschorsch2016cryptocurrency} a literature review on cryptocurrencies is presented, which is however not focused on BCT architectures. Moreover, in this work we focus not only on the cryptomarket but consider use-cases for blockchain other than cryptocurrency. The same issue recurs with other reviews focused on the literature related to smart contracts, i.e., programs that can be deployed and run on a blockchain \cite{BlockchainbMapping} and their applications \cite{macrinici2018smartapplications}. For example, Bartoletti and Pompianu \cite{bartoletti2017} focus on smart contract applications, and review the platforms and design patterns for such smart contracts. In our own work, we build from these foundations and include a comparison of smart contracts with the architecture principles of BCT.

Furthermore, there are several literature reviews that examine the use-cases and specific applications of BCT, e.g., Karafiloski et al. \cite{karafiloski2017blockchainbigdata}. Several other sector- and domain-specific reviews also exist, e.g., for the Internet of Things (IoT) \cite{internetofthingschristidis2016blockchains, khan2018iot}, or Multi-Agent Systems (MAS) \cite{calvaresi2018multiagent}. While other papers review how BCT could be utilized for the aforementioned domains, these reviews do not present BCT applications based on a rigorous scientific evaluation --- in our work we set out to operate a systematic synthesis of architecture elements as well as the alternatives in architecture decision-making spanning multiple domains and encompassing reference literature from, e.g., Supply-chain management \cite{korpela2017digital, wang2019understanding}, usage of BCT by governments \cite{batubara2018}, BCT in healthcare \cite{kuo2019comparison}, and more. 

By contrast, the work of Tama et al. \cite{tama2017critical} presents a brief critical review of BCT and some applications for multiple fields. Our work however, provides a more elaborate overview of BCT applications for these fields, and in addition an in-depth insight into the architecture of blockchain technology that has been obtained through a grounded theory approach. Furthermore, our study includes a definition of BCT based on formal concept analysis. What is more, Casino et al. \cite{casino2018systematic} present a review of BCT as a basis for multi-purpose applications design. Rather than concentrating solely on BCT applications, our work in addition provides a rigorous definition of blockchain based on formal concept analysis and offers an extensive multi-vocal catalogue and accompanying descriptions of anything that was published about blockchain technology. The catalogue presented in this paper follows the well-known 4+1 view framework \cite{kruchten19954+} for architecture description to aid anyone in framing, operating, deciding upon or describing blockchain architectures in general. Hence, the scope of our work is much broader, not only discussing applications of blockchain but in addition its architecture in depth. Finally, we provide a data-driven, in-depth, and evidence based overview of research gaps in the field of BCT.

\subsection{Structure of the Article}

The remainder of this paper is organized as follows: The next section reviews the background and basic notions of BCT. The methodology section (Sec. \ref{methodol}) elaborates on the approach taken to attain the results of this study. In the section thereafter (Sec. \ref{bctdef}) a definition of BCT that has been constructed based on the literature is presented. Following, the study presents BCT software architecture from multiple perspectives in \ref{bctarch}. In section \ref{scenarios} reviews the scenarios for using BCT found in the literature to provide insight into the applications for which BCT is used. The characteristics of BCT and architectural trade-offs are presented in section \ref{charactbct}. Challenges of BCT are presented in section \ref{challoutlook}, along with an outlook. The results of the research are discussed thereafter (Sec. \ref{discussion}). Based on the discussion of the results, in section \ref{gapsandfuture} suggestions for future research are presented. In section \ref{Limitations} the limitations of this research and potential threats to validity are addressed. Section \ref{conclusion} concludes the article.

\section{Background and Basic Notions}
\label{Basic Blockchain Technology Terminology}
As previously stated, BCT first appeared in 2008, featured in the seminal paper ``Bitcoin: A Peer-to-Peer Electronic Cash System" by Nakamoto \cite{nakamoto2008bitcoin}. The paper proposes a P2P electronic cash system that allowed for the execution of transactions between one party and another without requiring a trusted third party to act as a safeguard and check the validity of the transaction. A year later in 2009, the Bitcoin network was launched \cite{zohar2015bitcoin}. 

\subsection{Historical Setting}

The first solution that Nakamoto suggested to enable the transactions of digital coins is that owners of a coin wishing to commit a transaction should digitally sign a hash of the previous transaction and the public key of the next owner, both is added to the end of the coin. An electronic coin as such is defined as a chain of digital signatures. By verifying the signatures of a coin the payee can verify the historical chain of ownership. However, this provides a payee with no guarantee that the coin has not already been double spent as there is no way to verify that the previous owners did not sign any earlier transactions. \emph{Double-spending} refers to spending the same currency in two distinct transactions at the same time. In traditional settings, a centralized trusted third party (e.g. bank or mint) verifies whether the owner of a coin did not double spend the same coin. To verify transactions traditional trusted third parties maintain a centralized ledger which records all transactions and the order in which they were enacted. Moreover, the trusted third party needs to be aware of all transactions as there is no other way to confirm the absence of a transaction. 

\subsection{Terms and Definitions}

In order for transactions to be executed without a trusted third-party there also needs to be full awareness, and a single history of these transactions. In the Bitcoin paper two solutions are proposed to accomplish the aforementioned goals: (1) Transactions should be publicly announced to all participants in the network. These objectives are attained by employing a distributed ledger on a P2P network. Specific network participants called \emph{nodes} 
each store a local copy of the ledger. (2) Nodes need to reach a consensus about the history of the transactions, and the order in which they were received. This raises another problem however: Some of the nodes in the network might behave maliciously and try to change the communication contents. In literature this problem is referred to as the \emph{Byzantine Generals Problem} \cite{lamport1982byzantine}. Non-malicious nodes need to be able to distinguish the information that has been tampered with from the correct information by reaching a consensus over the consistency of the distributed ledger to determine the validity of a transaction. Consequently, this requires proof that when the transaction was executed, the majority of nodes have reached a trustworthy consensus that it was the first received. In essence, these requirements are introduced to ensure that the system to a certain extent can tolerate malicious behavior by nodes participating in the network, to which is commonly referred to as \emph{Byzantine fault tolerance} \cite{Peasebyzantinefaulttolerance1980}.    

In the seminal Bitcoin paper \cite{nakamoto2008bitcoin} several concepts are presented to satisfy these requirements starting with the use of a timestamp server. The server takes a hash of the block of transactions that are required to be timestamped and publishes the hash on the network. What the \emph{timestamp} proofs is that the data existed at a certain point in time. The hash is \emph{chained} to the previous hash because the latter time stamp is included in the former. As a result, each additional hash reinforces the ones before it, and as more blocks are added the chain will grow ever stronger. The next concept presented enables nodes to reach a consensus on whether the distributed ledgers are consistent with one another, thus that all transactions are valid. A naive way of  accomplishing this would be to let the majority of nodes vote over its consistency. However, that would make the blockchain prone to \emph{Sybil attacks} whereby a malicious attacker creates or copies multiple identities in order to control the network. 

Bitcoin diminishes the possibility of a Sybil attack by employing a \textit{Proof-of-Work} (PoW) \emph{consensus protocol} which stipulates that not the majority of IP-addresses count as the majority vote of the network, but rather the majority of computational power. While it might be easy for an attacker to create several nodes in a network, amassing large amounts of computational power might prove to be more difficult. The PoW consensus algorithm distributes accounting rights and rewards through a computing power competition in which all nodes of the network can participate. Nodes try to be the first to solve a computational hard mathematical puzzle by finding the right \emph{nonce} (a random number) for the block-header based on information of the prior block. This process is called \emph{mining} and the nodes executing the calculations are referred to as \emph{miners} in the Bitcoin nomenclature. The first miner to finish creates the next block and is rewarded by receiving an amount of Bitcoin. However, because the mining process is probabilistic two or more blocks might be created and propagated by distinct miners simultaneously. These phenomena are known as \emph{forks}. In the event of a fork, nodes as a rule always trust the \emph{longest chain} of blocks as the chain holding the truth with regard to transaction validity (which is analogue to the most computational work). Other nodes wait until new blocks are proposed after the occurrence of the fork to determine which chain will become the longest chain. Consequently, transactions are not confirmed before a longest chain has formed.

\emph{The longest chain rule} is a safeguard to secure the blockchain against the possibility to delay the propagation of transactions which in turn, opens the possibility of introducing fake transactions. As the computational power and interests of the miners might vary the PoW consensus protocol increases or decreases the difficulty of the mathematical problem in such a way that the interval between the generation of new blocks, referred to as \emph{block interval time} remains constant at 10 minutes. Tampering with the transactions recorded on the Bitcoin blockchain would therefore require an attacker not only to be the first one to generate the latest block, but also to control the longest chain.

After the introduction of the genesis PoW-based consensus protocol for Bitcoin many others have been introduced for blockchain such as: (1) \emph{Proof-of-Stake} (PoS), which replaces PoW based mining with a mechanism which makes the chances of mining a block proportional to the amount of stake (currency) a miner has \cite{li2017proof, swan2015blockchain, gatteschi2018blockchain, yu2018virtualization}; (2) \emph{Delegated-Proof-of-Stake} (DPoS), where the chances of mining a block are also based on a miner stake but allows for the delegation of voting on the correctness of a block \cite{mingxiao2017review, Zheng2017AnOverviewofBlockchainTechnologyArchitectureConsensusandFuturetrends}; (3) \emph{Proof-of-Elapsed-Time} (PoET) which used dedicated hardware to create consensus \cite{yaga2018blockchainoverview, meng2018intrusion}, and (4) \emph{Zero Knowledge Proofs} (ZKP) that aim to provide users performing transactions with more privacy \cite{yu2018virtualization,xu2017taxonomy, globalblockchain2017}. The advent of blockchain also revived the interest in preexisting consensus protocols such as \emph{Practical Byzantine Fault Tolerance} (PBFT), which could be utilized for a similar purpose as ZKPs \cite{dinh2018untangling, Zheng2017AnOverviewofBlockchainTechnologyArchitectureConsensusandFuturetrends, xu2017taxonomy}.

Furthermore, Bitcoin was envisioned as a \emph{public blockchain network} that anyone willing can access, and that is \emph{permissionless}, meaning that everyone connected to the network can request transactions or become a miner to check the validity of transactions. By contrast, in the past decade \emph{private blockchain networks} have been introduced that allow only selected participants from one organization to join the network which can also perform only actions that are \emph{permissioned} on the network. Finally, \emph{Consortium blockchain networks} can be considered a hybrid approach as the number of participants that can join the network is restricted, but they can be from different organizations. Among the connected participants the permissions they are granted on the network might differ \cite{Zheng2017AnOverviewofBlockchainTechnologyArchitectureConsensusandFuturetrends,xu2017taxonomy, pilkington2016}. 

Although the Bitcoin technology introduced the concept of BCT to allow for electronic payments using \emph{cryptocurrency} (digital coins) between anonymous peers, nowadays other blockchain networks such as Ethereum offer to possibility to deploy \emph{smart contracts}, that is, programs that can be deployed, run, and verified correct over a blockchain. Smart contracts use triggers, conditions and business logic to enable more complex programmable transactions \cite{xu2017taxonomy} for the automation of (business) processes \cite{gatteschi2018blockchain,frowis2017code,arevolutionintrust2017}.

\section{Research Methodology}
\label{methodol}
To attain our results, we conducted a systematic Multivocal Literature Review (MLR) on blockchain technology. Specifically, we address the following research questions:

\begin{enumerate}
\item[RQ1]{How can blockchain technology be systematically defined?}
\item[RQ2]{What applications of blockchain technology have currently been published and how can these applications be classified?}
\item[RQ3]{What are the properties of blockchain technology and what are their trade-offs?}
\item[RQ4]{What are the challenges for blockchain technology?}
\item[RQ5]{What are the current research gaps in the field of blockchain technology?}
\end{enumerate}

The first research question rotates around providing a systematically-derived definition for blockchains while RQ2 focuses on the applications for which BCT is utilized. RQ3 seeks to offer an overview of the notable properties of BCT (architectural or otherwise) as well as their trade-offs. The fourth research question focuses on delineating the challenges in the field of BCT. Finally, RQ5 aims at presenting research gaps that future research endeavors can address.

\subsection{Data Preparation Approach}
The benefit of a MLR approach is that, beyond typical systematic literature reviews \cite{kitchenham2004procedures} (SLRs) which use academic peer-reviewed articles alone, a multivocal literature review (MLR) also allows for the inclusion of Grey Literature (GL). GL is typically produced by practitioners, such as private industry, governments, academics and industry, and any party which is not controlled by commercial publishers or peer-review. Generally, therefore, grey literature is not published in books or scientific journals. However, this literature can provide invaluable insights into the state of the practice in a field \cite{garousi2017guidelines}. Given that at the moment of writing the field of BCT is still relatively in its infancy, we therefore deem it appropriate to include relevant literature created by practitioners in the field of BCT for a better understanding of the field. Including GL in our review allows us to combine and synthesize academic literature with the state-of-the-art in practice. 

\begin{figure}
    \centering
    \includegraphics[width=\textwidth]{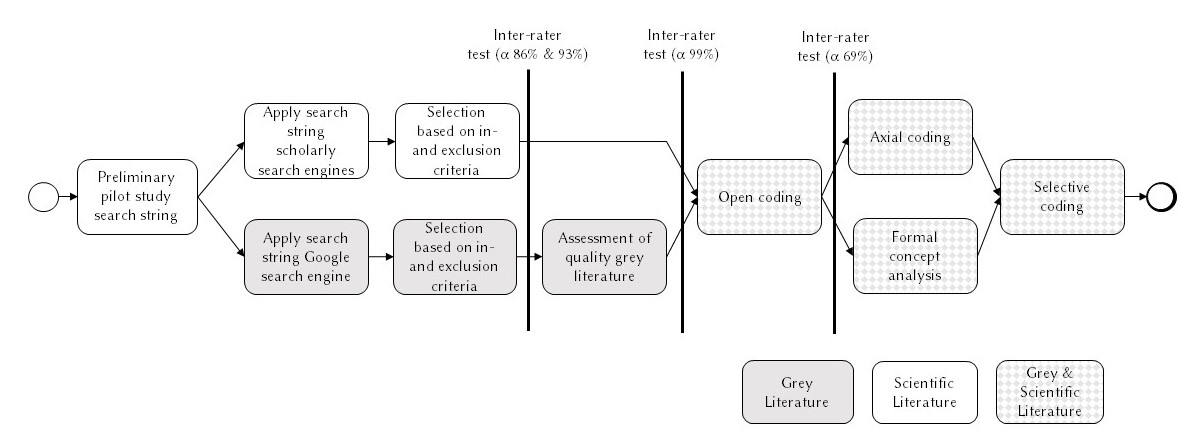}
    \caption{Research methodology, an outline.}
    \label{methodapp}
\end{figure}

In conducting our MLR we set out to identify (a) all relevant academic peer-reviewed articles (scientific literature), (b) all relevant grey literature for this study. To reduce the possibility of researcher bias, a predefined protocol for the identification of both the relevant scientific literature (SL) and grey literature (GL) needs to be established \cite{garousi2017guidelines}. While carrying out our systematic literature review we followed three steps: (1) Create a selection of articles to review. (2) Conduct the review (3) Analyze the data. A process model of the methodology used for this research is depicted in Figure \ref{methodapp}.

\subsection{Search strategy}
\label{Search strategy}

\begin{figure}
    \centering
    \includegraphics[width=\textwidth]{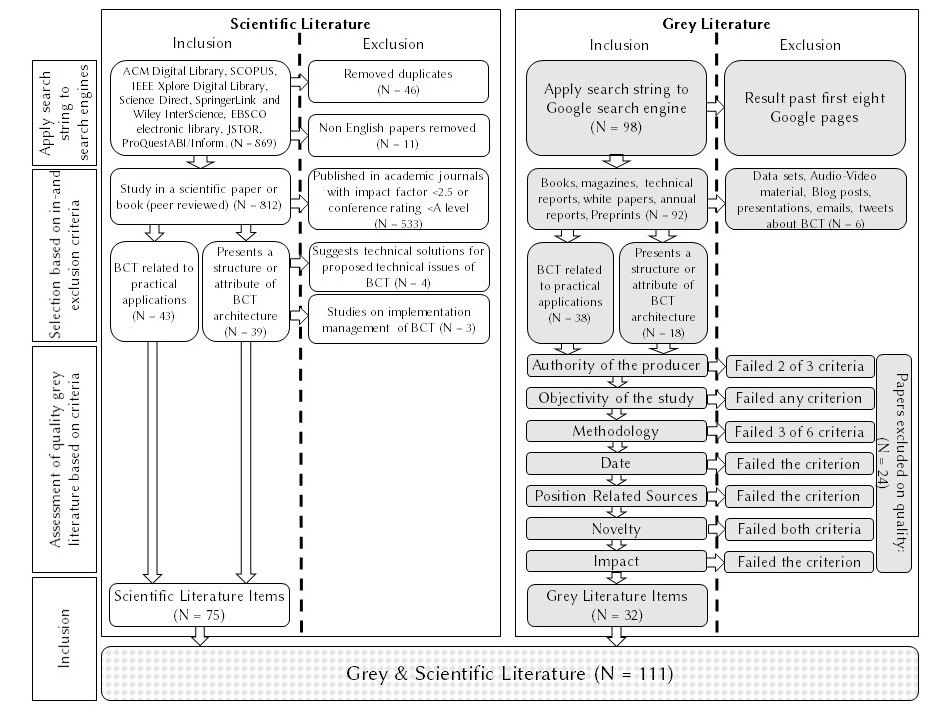}
    \caption{Sample Search and selection strategy, a process model.}
    \label{fig_searchstrategy}
\end{figure}

The first step has been carried out using the protocol for systematic reviews suggested by Kitchenham \cite{kitchenham2004procedures}. The protocol suggests three stages for a literature review: (1) elaborate the search string; (2) apply the string on chosen search engines; (3) Filter out and extract primary papers based on pre-established exclusion criteria from search results. The implementation of these steps is presented in Figure \ref{fig_searchstrategy}. 

The search string was determined by deriving relevant keywords from the research questions. Before carrying out our systematic search, we conducted a preliminary pilot study by experimenting with the search terms to select more results. This process yielded the following search terms:

(1)``Blockchain'' $\vee$ ``Blockchains'' $\vee$ ``Distributed'' $\vee$ ``Decentralized'' (2) ``Ledger'' $\vee$ ``Technology'' $\vee$ ``Database'' (3) ``Applications'' $\vee$ ``Use Case'' $\vee$ ``Implementation'' $\vee$ ``Example'' $\vee$ ``Case Study'' (4) ``Architectural'' $\vee$ ``Architecture'' $\vee$ ``Form'' $\vee$ ``Fabric'' $\vee$ ``Structure'' ``System'' $\vee$ ``Design'' (5) ``Choices'' $\vee$ ``Options'' $\vee$ ``Decisions''. When combined, the preceding terms were used in the following search string:
\begin{equation}
[(1\wedge2)\wedge3\vee(4\wedge5)]
\end{equation}

In the next stage (2), the search string has been applied to the following scholarly search engines: ACM Digital Library, SCOPUS, IEEE Xplore Digital Library, Science Direct, SpringerLink and Wiley InterScience, EBSCO electronic library, JSTOR knowledge storage and, ProQuestABI/Inform throughout March in 2018. The final stage (3) of the systematic review, the initial results were screened against inclusion and exclusion criteria that are shown in Fig. \ref{fig_searchstrategy} (Selection based on in-and-exclusion criteria)\footnote{The (N = followed by a number) in Fig 2. represents the number of papers included or excluded based on the selection criteria.}. For brevity's sake we have not included a more elaborate description and rationale behind these criteria in the paper, but they can be accessed online (see \ref{replpack} for more details).

For the second part of the MLR, to identify all relevant Grey literature, we established another protocol to filter and extract the GL using the guidelines suggested by Garousi et al. \cite{garousi2017guidelines}. The protocol has been conducted in three stages: (1) Search process, (2) Source selection, (3) Study quality assessment. The implementation of these steps can be found on the right-and side of Fig. \ref{fig_searchstrategy}. 

In the first stage, we applied the search string to the Google search engine. The search process yielded 8.330.000 results when applying the first search string (``Blockchain'' $\vee$ ``Ledger'' $\vee$ ``Applications''). Because of the significant amount of results we initially limited our review to the first eight pages (20 results per page) provided by the Google search engine. Incrementally the next pages thereafter have been reviewed using inclusion and exclusion criteria related to the type of grey literature source (e.g. books, magazines or video files), (see Fig. \ref{fig_searchstrategy}). Thereafter the pages were incrementally reviewed by title and abstract, starting from the first results page using the inclusion and exclusion criteria depicted in Fig. \ref{fig_searchstrategy} (selection based on in-and exclusion criteria). If <50\% of the results on a page were not relevant for this research, the search was stopped there. We further refined the GL studies we obtained from the first eight Google pages using the same inclusion and exclusion criteria.

During the second phase, we assessed the quality and relevance of the sources of the primary GL we obtained since it cannot be assumed that the quality of GL is guaranteed. Exclusion criteria suggested by Garousi et al. \cite{garousi2017guidelines} have been used for this purpose (see Fig. \ref{fig_searchstrategy}). 

The exclusion criteria used consist of 7 quality categories ranging from the authority of the producer to the objectivity of the study that can be found in Fig. \ref{fig_searchstrategy} under assessment of quality grey literature. Combined these quality categories encompass 17 criteria that have been assessed one by one for each GL item. An overview of all the quality categories, quality criteria, and how many of these criteria had to be satisfied to include the item can be found in Table \ref{qualitygl}. 

\begin{table}
\caption{Quality criteria grey literature.}
\label{qualitygl}
\resizebox{\textwidth}{!}{%
\begin{tabular}{@{}lll@{}}
\toprule
\textbf{Category}                  & \textbf{Exclusion Criteria}                                                                                                                                       & \textbf{Criteria to Satisfy}     \\ \midrule
Authority of the producer & \begin{tabular}[c]{@{}l@{}}The publishing organization is reputable, or the individual author\\ is associated with a reputable organization\end{tabular} & \multicolumn{1}{c}{2/3} \\
                          & The author has published other work in the field                                                                                                         &                         \\
                          & The author has expertise in the area (e.g. job title)                                                                                                    &                         \\
Objectivity of the study  & The statement of the sources is objective                                                                                                                & \multicolumn{1}{c}{3/3} \\
                          & There are no vested interests                                                                                                                            &                         \\
                          & Conclusions are supported by data                                                                                                                        &                         \\
Methodology               & The source has a clearly stated aim                                                                                                                      & \multicolumn{1}{c}{4/6} \\
                          & The source has a clearly stated methodology                                                                                                              &                         \\
                          & The source is supported by authoritative, documented references                                                                                          &                         \\
                          & Limits are clearly stated                                                                                                                                &                         \\
                          & The work covers a specific question                                                                                                                      &                         \\
                          & The work refers to a particular population                                                                                                               &                         \\
Date                      & The item has a clearly stated date                                                                                                                       & \multicolumn{1}{c}{1/1} \\
Position related sources  & Key related GL or formal sources have been linked/discussed                                                                                              & \multicolumn{1}{c}{1/1} \\
Novelty                   & The item enriches or adds something unique to the research                                                                                               & \multicolumn{1}{c}{1/2} \\
                          & The item strengthens or refutes a current position                                                                                                       &                         \\
Impact                    & \begin{tabular}[c]{@{}l@{}}The GL source should have citations and backlinks\\ to substantiate the arguments made in the study\end{tabular}              & \multicolumn{1}{c}{1/1} \\ \bottomrule
\end{tabular}%
}
\end{table}

The authors of this study (viz., the first two authors of this study), have indicated whether a GL item: (a) satisfied, (b) did not satisfy a criterion. In the cases where one of the criteria could not be assessed (e.g. because this information was missing) we have assessed these criteria as if they did not satisfy the criterion. GL items that did not satisfy the threshold for each quality category were excluded from the sample. After the selection process we merged the grey-and scientific literature into one sample as the literature under review.

\subsection{Data Analysis}
\label{Data Analysis}
This section details the analysis methods enacted to address our research questions.

\subsubsection{Formal Concept Analysis}

To address RQ1, a Formal Concept Analysis (FCA) approach was adopted. FCA is a systematic approach to derive a formal ontology or concept hierarchy from a set of objects and their attributes \cite{vskopljanac2014formal}. A complete description of the FCA method employed to attain our definition of BCT can be found in Appendix \ref{app_FCA}.

\subsubsection{Grounded-Theory Analysis}

For step 2 and 3 of the MLR, and to address RQs 2,3 and 4 a Straussian Grounded Theory (GT) approach \cite{glaser2017discovery} was adopted. In the scope of straussian GT, a series of systematic steps are enacted to allow a theory to emerge from the data (hence, ``grounded'') using codes. For our research each code represents a concept or theme related to BCT. Whenever a paragraph in the literature under review represented one of these concepts or themes, the related appropriate code has been attached. In GT this process known as "coding", and includes the phases that are described in Appendix \ref{app_GT}. 

\subsection{Inter-Rater Reliability Assessment}
\label{interrater}
We employed Krippendorff coefficient (or K-$\alpha$) \cite{hayes2007answering}, to evaluate the inter-rater reliability of the inclusion and exclusion of SL items, the in- and exclusion of GL items, the quality assessment of the GL, and the coding process of the pilot study. The coefficient measures the agreement between two ordered lists of codes which have been applied as part of content analysis. The methods used to asses the inter-rater reliability between the raters, and the results thereof can be found in Appendix \ref{app_interrater}.

\subsection{Sample Selection Results}
\label{sec: results}

This section outlines the sample results of our search strategy (Sec. \ref{Search strategy}). First the section presents the distribution of the sample between grey and scientific literature, along with the distribution the publication venues per year. Thereafter, the section showcases a frequency analysis of the topics discussed in the papers under review along with an overview of these trends per year. Finally, the distribution of these topics between grey and scientific items is presented and discussed.

\subsubsection{Publication Venues and Distribution Literature}

\begin{figure}
    \centering
    \includegraphics[width=\textwidth]{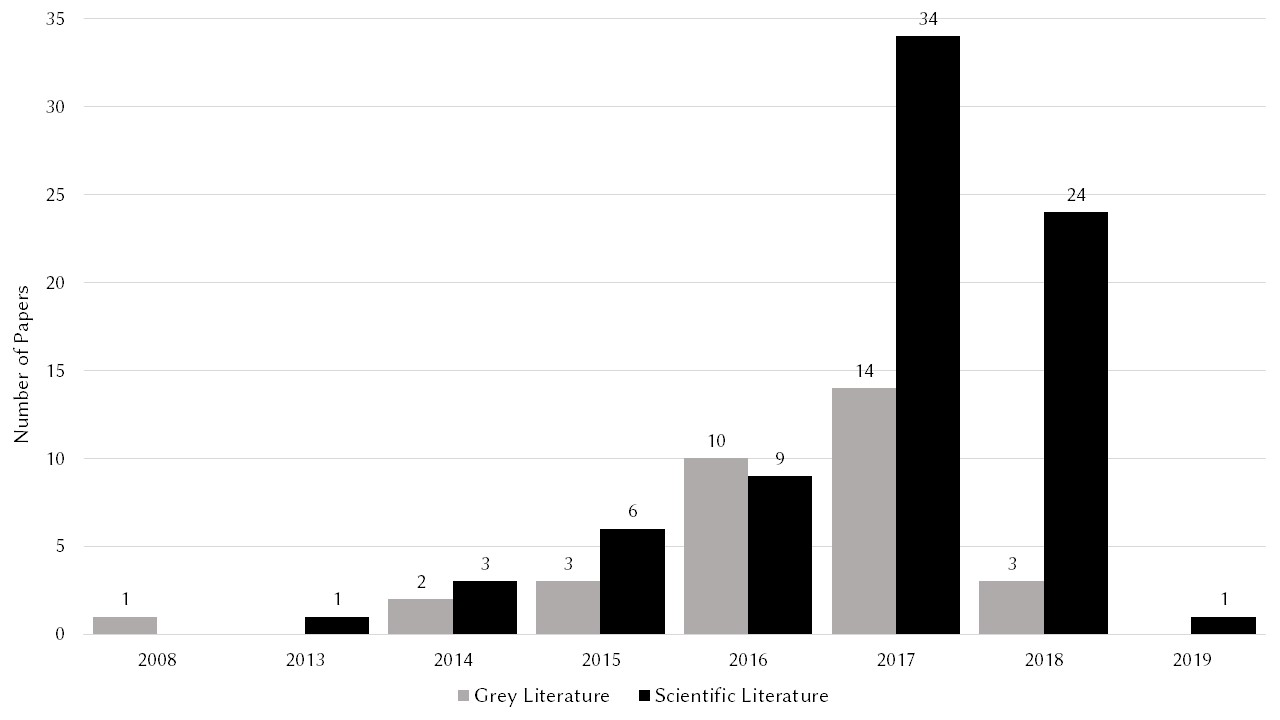}
    \caption{Sample results; Grey and Scientific Literature across primary studies.}
    \label{stats_gandw}
\end{figure}

After the assessment of the GL, the search strategy yielded a total of 33 GL items and 78 of scientific peer-reviewed papers selected for this study. From this point on, we no longer distinguish the results whether they were derived from GL or SL but flesh out results over the total of 111 studies be the object of this research. Figure \ref{stats_gandw} depicts the number of SL and GL per publication year. The results depicted for the year 2018 have only been collected until March 2018 and therefore might skew the results.

\begin{figure}
    \centering
    \includegraphics[width=\textwidth]{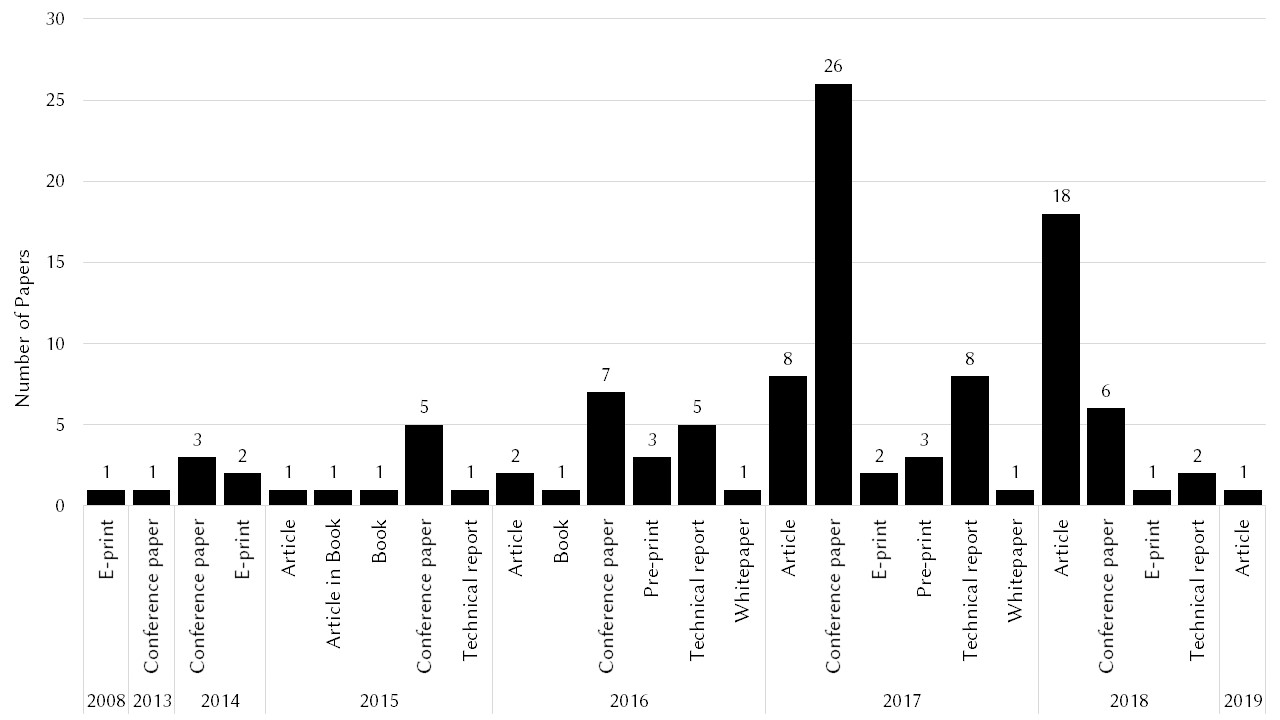}
    \caption{Sample results; Publication Venues per Year.}
    \label{stats_venuepub}
\end{figure}

Figure \ref{stats_gandw} shows that from 2008 until 2013 BCT has gained little attention from either practitioners or scholars. The figure show an overall increase in the SL and GL published from 2014 onwards. Furthermore, the statistics show that there is a growing interest from the scientific community for research in BCT. More specifically, from 2017 onwards twice as many articles have been published as compared to the years before. However, these results also indicate that research in the field of BCT is still in its infancy given that from 2008 to 2016 little scientific work has been published. The search and selection results also indicate an increase in the amount of practitioners literature being published. 
Furthermore, the sources from which these research items were identified for this study are diverse (see Fig. \ref{stats_venuepub}), ranging from articles in technical magazines, books, and technical reports alike. The majority of items however, were published in conference proceedings and reflect white literature. 

In the years directly following 2008, i.e., the introduction of BCT, publications on the topic were almost evenly distributed among different sources (e.g. books, conference papers). However, as of 2016, books on BCT have not been found by this research. Although the publication of conference papers shows an increase in 2017, articles published in scientific venues are more gradually increasing in frequency.

\subsubsection{Topics in the Field of BCT}
\label{mainthemes}

\begin{figure}
    \centering
    \includegraphics[width=\textwidth]{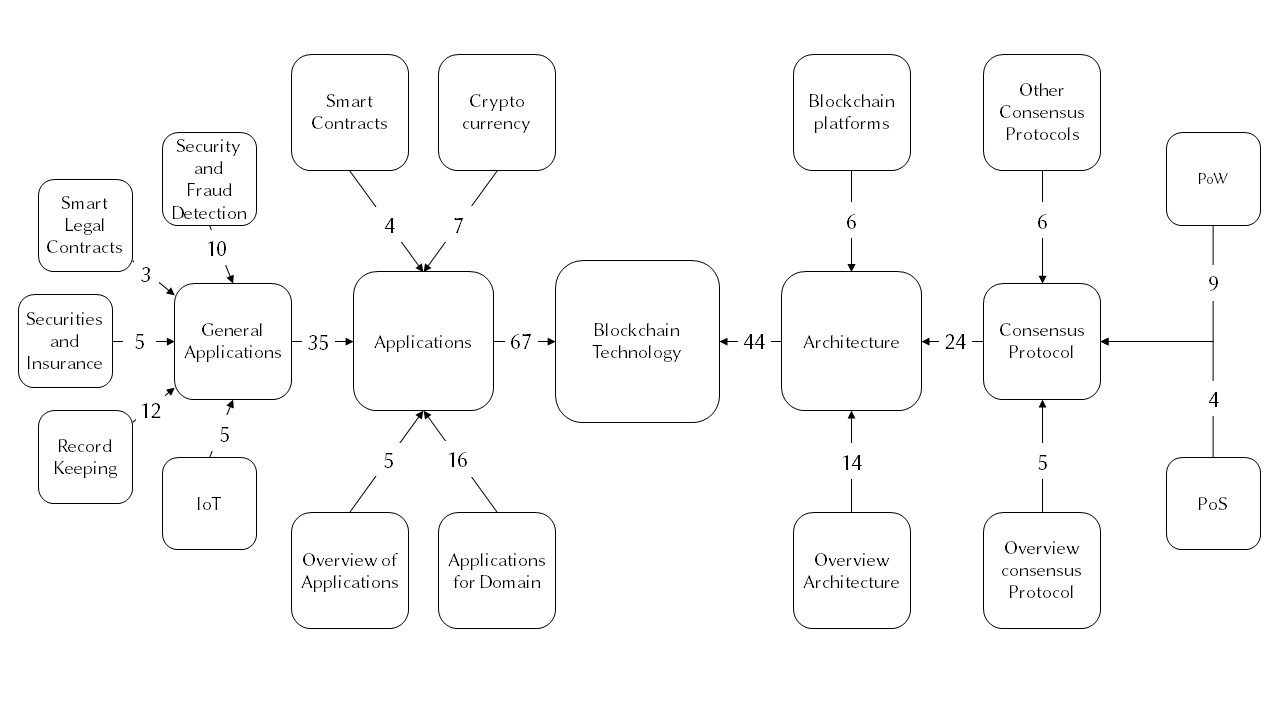}
    \caption{Sample results; Topics frequency analysis.}
    \label{fig_maintopics}
\end{figure}

Fig. \ref{fig_maintopics} shows the the main topics of the papers under review, as elicited using a grounded-theory approach. More specifically, the blocks in Fig. \ref{fig_maintopics} represent the topics found in the literature, while the number on the arrows between the blocks represents the \emph{weight} of the topic, in terms of number of papers where those topics were coded. The direction of the arrow itself depicts under which of the composed main topics the sub topics are categorized. Among the items selected for this study, BCT-based applications are strongly represented (see left-hand side of Fig. \ref{fig_maintopics}) while items on BCT architecture and are represented slightly less, with a ratio of 2/3. 

In terms of applications, five sub-themes can be distilled: (1) cryptocurrencies, (2) smart contracts, (3) papers that provide an overview of BCT, (4) literature that suggests applications of BCT for specific domains, and (5) finally, literature that presents general applications for BCT. Interestingly, Fig. \ref{fig_maintopics} again shows that the applications of BCT for a specific domain and general applications of BCT has gained considerable attention. 
On the one hand, literature on BCT for specific domains encompasses complex domains such as e-government, the financial sector, and relief development \cite{chow831}, while literature on the general topic of smart contracts is rather limited. 
On the other hand, the specific technical architecture literature over BCT reflects five categories: (1) security and fraud detection; (2) smart legal contracts; (3) securities and insurance; (4) record-keeping; (5) the Internet of Things (IoT). Among these categories the majority of papers has been published on utilizing BCT for record-keeping, that is, registering certain data on the blockchain to ensure its immutability. Another major focus of research on BCT applications is security and fraud detection, these items focus on using blockchain for safe distribution of data among peers. A more detailed description of the contents of literature of applications for BCT can be found in section \ref{scenarios}.

\subsubsection{Trends in Publications on BCT}

\begin{figure}
    \centering
    \includegraphics[width=\textwidth]{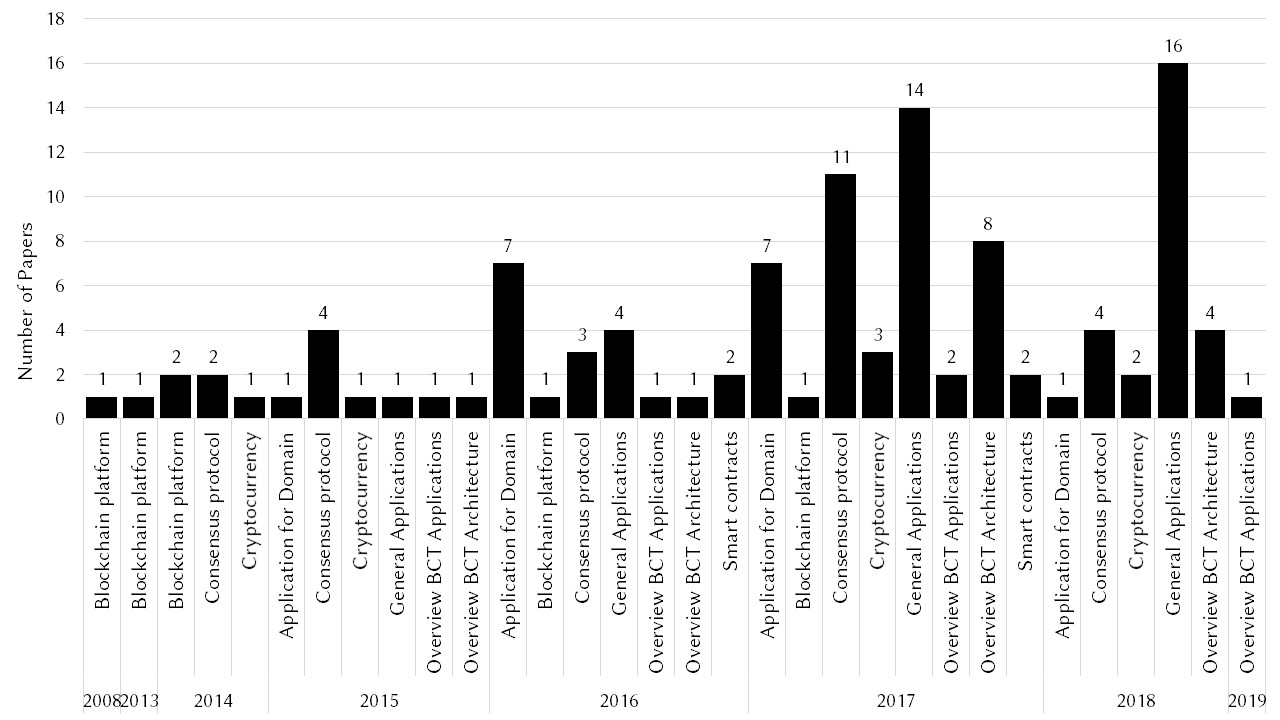}
    \caption{Trends in BCT publication topics.}
    \label{fig_trendstopics}
\end{figure}

Four key trends in the literature under review can be observed (see Fig. \ref{fig_trendstopics}). First, there is a balance in the distribution among topics even though the overall number of works on BCT is steadily increasing.

Second, two exceptions are (a) works related to applications for specific domains and (b) general applications research, as previously discussed --- for these, the years 2016 to 2018 have seen a tremendous increase in research and practical work. 

Third, publications on blockchain-based smart contracts have seen a rise only from 2016 onwards. A similar observation can be made for items related to cryptocurrency as only one paper was published in 2015, with 3 papers being published in 2018.

Lastly, Since 2014 an increasing amount of papers on the topic of consensus protocols have been published, a trend that continues to date.

\begin{figure}
    \centering
    \includegraphics[width=\textwidth]{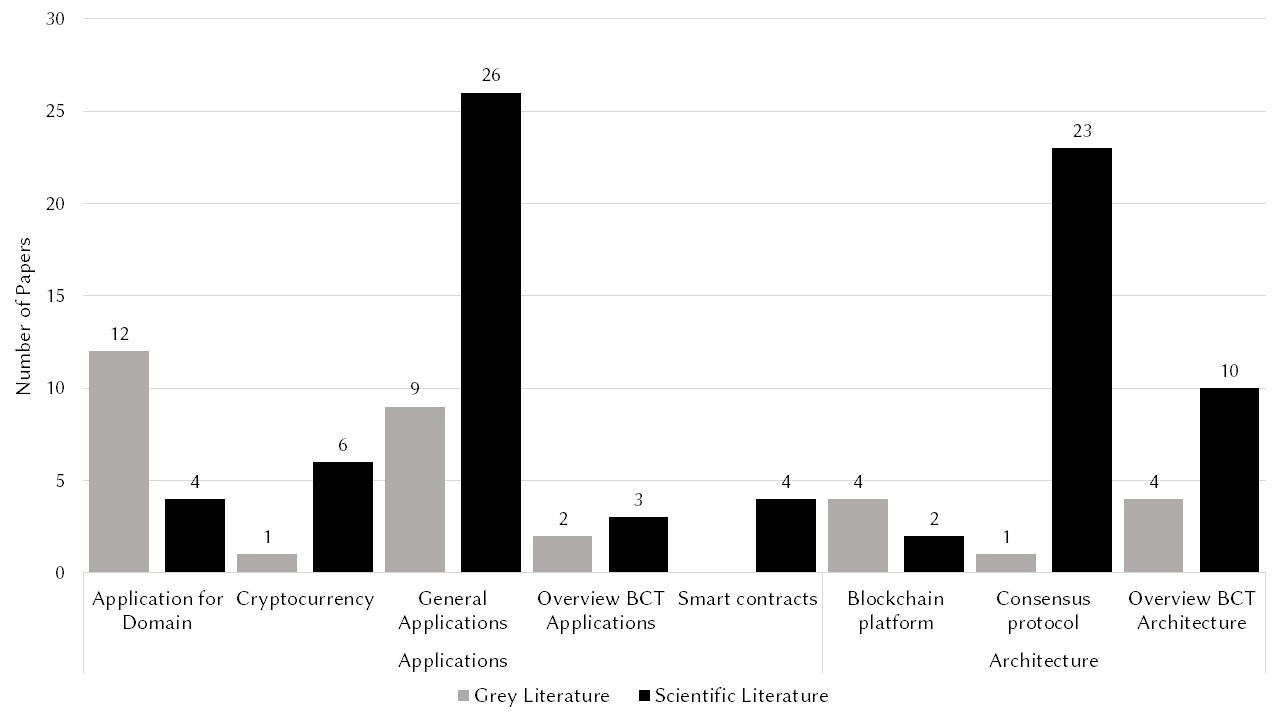}
    \caption{Distribution topics scientific and grey literature.}
    \label{stats_sciencegreythemes}
\end{figure}

What is more, in terms of past publications by both practitioners and scholars have predominantly been focused on applications for BCT (see Fig. \ref{stats_sciencegreythemes}) with more of the grey literature being published on applications (72\%) as compared to scientific literature (55\%). 

With respect to engineering research focus, scholars have focused their efforts on consensus protocols whereas practitioners have mostly presented works on blockchain platforms or overviews of BCT architecture.

\section{A Systematic Definition of Blockchain Technology}
\label{bctdef}
So far we have referred to BCT as a single technology, however BCT is a clever combination of several technologies and elements and there is no consensus on the definition of a blockchain \cite{globalblockchain2017}, with a precise definition of blockchain technology often subject to controversial and subjective opinion \cite{applicationsofdistributed2017}. Stemming from the data available to us, We strived to construct a rigorous definition of BCT based on (a) the identified software elements drawn from literature, (b) relations among them and (c) their properties\footnote{Public-key cryptography is the commonly used to describe the exchange of information using a set of private and public keys. Hence, when constructing a definition the term public-key cryptography has been used instead of private and public keys as an attribute of BCT.} reflecting the software architecture research and practice state of the art \cite{bassSoftwareArchInPractice}.

First, our data indicates that: \emph{Blockchain technology} is a form of \emph{distributed ledger technology}, deployed on a \emph{peer-to-peer network} where all data is replicated, shared, and synchronously spread across multiple peers. The technology allows actors participating in the network to perform, sign, and announce \emph{transactions} by employing \emph{public key cryptography}. Transactions are executed following a \emph{consensus protocol} operated by specific nodes to ensure the validity of transactions requested by other peers in the network, and to synchronize all shared copies of the distributed ledger. During a consensus protocol execution, the data of valid transactions, along with other required metadata concerning the network, and the hash of the previous block is bundled into a \emph{block} using \emph{hashing functions}. The essential and key property reflecting BCT architectures is that each block contains the hash of their predecessor, therefore linking all prior transactions to newly appended transactions; the blocks therefore form a \emph{chain} with the aim of establishing a tamper-proof historical record.

\section{Blockchain Technology: Architecture Elements}
\label{bctarch}

This study examines the blockchain architecture landscape, arranging the elements found in literature through the well-known 4+1 software architecture framework introduced by Kruchten \cite{kruchten19954+}. The Kruchten framework delineates the comprehensive interplay of relations, properties and software elements in BCT and encompasses five views, namely: (1) logical view; (2) development view; (3) process view; (4) physical view; (5) a use case view. In Appendix \ref{app_4+1} a more elaborate description of these views is provided.

Using the logical view, we first delineate the architectural elements to present the functionalities that various end users ultimately use from a blockchain. Further on, the development view describes how building BCT can be divided into smaller chunks of programmable code. Subsequently, the process view shows how IT systems behave during run time and is of interest to system integrators that need to know about the thread of control to execute operations utilizing BCT. Beyond that, the physical view is of interest to system engineers that maintain overall blockchain system, also, given that BCT completely resides on a P2P network its different arrangements is discussed in the physical view. The use-case view of the 4+1 model is recapped later in Sec. \ref{scenarios}.

\subsection{Logical View}
\label{logicalview}

The logical view emphasizes on the functional requirements and services the system should provide to its end users \cite{kruchten19954+}. Decomposition of the architecture aids in identifying the elements that are common across the system. We used an ontology for BCT as proposed in \cite{de2017understanding, glaser2017pervasive} to organize the discussion of its main elements. A more elaborate, and in-depth description of BCT elements is provided in Appendix \ref{app_logicalview}.

Blockchains are transaction oriented. Transactions in a blockchain system are executed using \emph{public key cryptography}. \emph{Cryptographic Hash functions} are used for the purpose of many operations, such as signing transactions (SHA-256 in the Bitcoin case\cite{nakamoto2008bitcoin}). The peers in the P2P network, also referred to as nodes are devices capable of processing and verifying transactions. Depending on the permissions all nodes or a specific subset of nodes validate transactions. The \emph{permissions}. There exists at least three categories of blockchain networks \cite{xu2017taxonomy,Zheng2017AnOverviewofBlockchainTechnologyArchitectureConsensusandFuturetrends}; Public, private, and consortium networks that have different arrangements in terms of their permissions. On a blockchain transactions are stored in \emph{blocks}. Each block is linked to its predecessor known as \emph{parent block} by including its blockheader hash to form an integral chain of blocks that can be traced back to the first, or \emph{genesis block}. Hence the term "blockchain" technology. Novel blocks are generated using a consensus protocol. Provided that the transactions included in the newly proposed block are valid, each new block enhances the security guarantees of the block before it \cite{DistributedLedgerTechnologyandBlockchain2017,nakamoto2008bitcoin,mingxiao2017review}. Updates and changes to the software of a blockchain are called \emph{forks}.

\subsection{Development View}
\label{developmentview}
Existing blockchain networks can be leveraged to build \emph{decentralized applications} (DApps) upon that use their services. Developers seeking to build their own blockchain platform have to program multiple software packages. Enabling transactions forms the basis for any blockchain network. A wallet needs to be programmed to allow clients of the platform to interact with other peers in the network. An \emph{address propagation} method should be installed for nodes to interact. Next the nodes need to connect via \emph{peer discovery}. Another aspect is the mechanism for \emph{propagating transactions} \cite{biryukov2014deanonymisation}. 

Data with regard to transactions can be stored in two ways: As a first method, like the Bitcoin, one can choose to add data into transactions. Another second method is to add data into contract storage like Ethereum \cite{xu2016connectblockchain}. Finally, an existing consensus protocol can be selected to process transactions or the protocol can be designed from scratch. Appendix \ref{app_developmentview} further describes the development view. 

\subsection{Process View} 
\label{processview}

The process view specifies which thread of control execute the operations of the classes identified in the logical view. The consensus protocol is at the heart of all BCT processes since it allows for the enactment of transactions and ensures that the distributed ledger remains consistent. Appendix \ref{app_processview} delineates the steps, issues and potential variants of the consensus protocols discussed in this section more in detail. 

\subsubsection{Practical Byzantine Fault Tolerance}(PBFT).
PBFT is mostly used in a private setting for permissioned blockchains because it assumes authenticated nodes \cite{dinh2018untangling, Zheng2017AnOverviewofBlockchainTechnologyArchitectureConsensusandFuturetrends, xu2017taxonomy}. The protocol itself is exclusively based on communication, and nodes go engage in multiple rounds of communication to reach consensus \cite{dinh2018untangling}. Nodes do not get a reward for achieving consensus, rather in the event of malicious behavior by an authenticated node it can be held legally accountable \cite{peters20166understanding, globalblockchain2017}. A primary leader node mines the blocks. The leader can be changed by other nodes via a "view-change" voting protocol, in the occurrence of a crash or when it exhibits malicious behavior \cite{internetofthingschristidis2016blockchains, mingxiao2017review}.

\subsubsection{Proof-of-Work}\label{pow}(PoW) is often referred to as the \emph{Nakamoto consensus protocol} \cite{luu2016making, voyiatzis2017merged, luu2015demystifying,pass2017sleepy}. The PoW consensus protocol is designed for the case where there is little to no trust amongst users of the system \cite{yaga2018blockchainoverview}. Public blockchains need to have a high degree of Byzantine fault tolerance as users can not trust one another.

Consensus in PoW is achieved through a hashing competition between miners. Competing miners need to commit computing power to calculate the solution to the same mathematical problem. To incentivize miners to participate in the consensus process the miner that is the first to find the solution to the mathematical problem reserves the right to publish the next block, and is rewarded by an amount of cryptocurrency \cite{howblockchain, yaga2018blockchainoverview, xu2017taxonomy, mougayar2016business}. In addition, the miner to win the competition with its peers is also be able to collect the transactions fees that were paid by clients.

Finding the solution to a PoW problem is a computationally arduous process for which there are no shortcuts \cite{DistributedLedgerTechnologyandBlockchain2017, yaga2018blockchainoverview}. The solution to the problem is hard to find, yet easy to check once they have been found \cite{luu2015demystifying}. Given that only one miner can win the competition and is rewarded the other nodes have simply wasted resources (CPU power and energy) in their attempt \cite{swan2015blockchain,yaga2018blockchainoverview,gatteschi2018blockchain,mingxiao2017review,xu2017taxonomy}. In addition, because the difficulty of PoW problems increases over time makes it even harder to win the competition \cite{yaga2018blockchainoverview}.

\subsubsection{Proof-of-Elapsed Time}(PoET) PoET is designed to address the inefficiency of PoW and replaces it with a protocol that is based on \emph{trusted hardware}. A node that uses trusted hardware however, can be checked for certain properties such as whether it is running a certain software. This aids in relaxing the trust model in settings were the Byzantine's Generals Problem might be present \cite{dinh2018untangling}. Sawtooth Lake, a project by Hyperledger, leverages Intel's \emph{Software Guard Extensions} (SGX) to establish a validation lottery that makes use of their CPUs capability to render a timestamp that is cryptographically singed by the hardware \cite{DistributedLedgerTechnologyCybersecurity2016}.

\subsubsection{Proof-of-Stake}(PoS) \label{PoS}As a response to the limitations of PoW the BCT community has turned towards Proof-of-Stake (PoS). The PoS consensus protocol has been introduced for public settings \cite{mingxiao2017review} with the aim to safeguard against Sybil attacks and malicious behavior by untrusted nodes \cite{dinh2018untangling}. The PoS protocol offers a more efficient and environmental friendly alternative to PoW as computing power is partially substituted by virtual resources (e.g. cryptocurrencies) that miners must invest to propose blocks \cite{li2017proof, swan2015blockchain, gatteschi2018blockchain, yu2018virtualization}. Rather than using computer power as a scarce resource to generate security, Proof of Stake uses the scarcity of the coin itself. Therefore nodes that participate in a PoS consensus protocol are more commonly referred to as \emph{forgers} instead of miners \cite{baldimtsi2016indistinguishable, li2017securing}. 

The idea behind the PoS model is that the more assets (e.g. cryptocurrency), or \emph{stake} a node has, its incentive to undermine the system diminishes because subverting the system would inherently mean that the worth of the nodes' stake would decrease \cite{yaga2018blockchainoverview, meng2018intrusion}. Logically, this implies that one cannot participate in the consensus protocol without owning a stake \cite{gatteschi2018blockchain}. A shared commonality of all PoS variants is that nodes that have more stake have a higher chance of generating new blocks \cite{yaga2018blockchainoverview, li2017proof, mingxiao2017review, Zheng2017AnOverviewofBlockchainTechnologyArchitectureConsensusandFuturetrends}. In other words, the more skin a forger puts in the game the higher its reward will be.

\subsubsection{Delegated-Proof-of-Stake}(DPoS) Delegated Proof-of-Stake introduces another variant of PoS \cite{mingxiao2017review, Zheng2017AnOverviewofBlockchainTechnologyArchitectureConsensusandFuturetrends}. In DPoS stakeholders elect delegates, referred to as \emph{witnesses} to forge and validate blocks in round-robin fashion \cite{li2017securing}.

Compared to PoW and Pos, DPoS is more energy efficient. Further, because the voting about the validity of a block is delegated and fewer nodes are needed to validate the blocks can be confirmed more quickly. Hence, as compared to PoW and PoS, DPoS has a low latency. Moreover, parameters including block size and block intervals can be adjusted by \emph{committee members} of the governance board. When a delegate acts malicious this dishonest delegate can be voted out by all the other nodes \cite{Zheng2017AnOverviewofBlockchainTechnologyArchitectureConsensusandFuturetrends,li2017securing}.

\subsubsection{Zero-Knowledge-Proofs}
\label{zkp}
Recently, different Zero-Knowledge-Proofs (ZKP's) based BCT networks have been proposed to preserve users' anonymity and confidentiality of transactions \cite{xu2017taxonomy}. In general, ZKP's aim to confirm a statement about a transaction such as "This is a valid transaction" without revealing anything about the transfer (statement) itself or the parties involved \cite{yu2018virtualization,xu2017taxonomy, globalblockchain2017}. Zerocoin was the first initiative with the aim of providing transaction unlinkability using ZKP's \cite{dinh2018untangling}. Similar to the Bitcoin Zerocoin uses the PoW consensus protocol to validate transactions. A \emph{cryptographic mixer} is implemented for Zerocoin to conceal the links between a zerocoin and the corresponding Bitcoin.

Building on the ZKP approach as a foundation, Zcash, extent the privacy guarantees, and improve the efficiency (throughput and latency) of Zerocoin. Zcash uses a variant of the PoW called Equihash. Transactions made using Zcash, including the split and merge transactions, are fully private \cite{dinh2018untangling}. Zcash employs a technique called \emph{Zero-Knowledge-Succinct Non-Interactive Argument of Knowledge} (zk-SNARKS) to provide these privacy guarantees \cite{yu2018virtualization, mingxiao2017review} that are a specific type of ZKP.

\subsection{Physical View}
\label{physicalview}
The physical view is concerned with the topology of software components and their physical connections. Electronic devices known as nodes constitute to a blockchains' P2P network and are the only physical connection to the non-digital world. P2P networks on which blockchain platforms are run have different arrangements; First, the network can be categorized on the basis of
permissions (authorization). Second, networks can be categorized with regard to their accessibility. Permissions to perform operations on the blockchain might differ ranging from allowing anyone to read, write and to partake in the consensus protocol to only one of these permissions. Control over these permissions can be confined to a distinct group of nodes, or all nodes.

As the name suggest \textit{Permissionless} grant permission to all nodes in the P2P network to read and write transactions. \emph{Permissioned} blockchain platforms have confined and idiosyncratic permissions for their nodes \cite{xu2017taxonomy, yaga2018blockchainoverview, distributedledgertechnologyinpayments2016, arevolutionintrust2017}.

The P2P network can also be described from the perspective of network accessibility. In the literature three categories of P2P networks can be distinguished that are coupled to a permission model \cite{Zheng2017AnOverviewofBlockchainTechnologyArchitectureConsensusandFuturetrends,xu2017taxonomy,globalblockchain2017,deblockchainsforgovernmental2017distributed,de2017understanding,cachin2017blockchains}. A \emph{Public blockchain}, like the Bitcoin or Ethereum have \emph{open network access} meaning that anyone willing is allowed to join the network. \emph{Private blockchains} are blockchains networks that are owned by one organization. Contrairy to public blockchains access is confined. \emph{Consortium blockchains} are similar to private blockchains in the sense that nodes first need to be authenticated before granted access to the network. However, consortium blockchains allow nodes from different organizations to access the blockchain network \cite{xu2017taxonomy, mingxiao2017review,pilkington2016}. A more elaborate description of both models and blockchain networks can be found in Appendix \ref{app_physicalview}. 

\section{Blockchain Use-Case View: Main Usage Scenarios}

\label{scenarios}

The use-case view aims at providing a description of an architecture by illustrating an essential set of use cases and scenarios for their usage. Our data suggests there are mainly three flavours of BCT applications: (1) Cryptocurrencies, (2) Smart contracts and (3), general-purpose applications. Following these versions we further categorize BCT applications \cite{swan2015blockchain}. For instance, the general-purpose applications of BCT can be arranged into five additional categories that encompass: (1) Security and Fraud Detection, (2) Securities and Insurance, (3) Record-Keeping, (4) Internet-of-Things, (5) Smart Legal Contracts. A complete catalogue of BCT applications can be found in Appendix \ref{app_usecasecat}.

\section{Blockchain Technology: Main Architecture Properties}
\label{charactbct}
The scope of our analysis revealed 8 essential architectural properties with a directed mutual influence relation evident from the state of the art. In fact, stemming from the relations our GT analysis we marked with a $\leadsto$ operator the mutual implication relation evident between the following couples of properties:
\begin{itemize}

\item \textit{Decentralization $\leadsto$ Disintermediation}. In traditional centralized transaction systems each transaction needs to be validated by a (trusted) third party (e.g., a bank). The decentralized workings of BCT enables the direct transfers of digital assets between two counter parties without this third party leading to direct disintermediation \cite{arevolutionintrust2017,DistributedLedgerTechnologyandBlockchain2017,gatteschi2018blockchain,Zheng2017AnOverviewofBlockchainTechnologyArchitectureConsensusandFuturetrends}. 
\item \textit{Programmability $\leadsto$ Automation}. BCT allows for the execution of pre-defined conditions that are automatically executed once certain conditions have been met. BCT enabled smart contracts extend this concept further by allowing (Turing complete) programmability of transactions
\cite{arevolutionintrust2017,DistributedLedgerTechnologyCybersecurity2016,clack2016smartfoundations}. The Bitcoin blockchain predominantly offers a service to exchange cryptocurrency and, accordingly provides limited support for smart contracts Blockchains like Ethereum or Kadena offer a fully programmable smart contract environment \cite{arevolutionintrust2017,DistributedLedgerTechnologyandBlockchain2017,globalblockchain2017,meng2018intrusion}. Offering smart contracts as a service however, adds another layer of complexity; smart contract execution puts a higher strain on the data storage requirements, throughput and latency of a blockchain network \cite{globalblockchain2017, de2017understanding, EthereumHomesteadDocumentation2017}. Furthermore, arbitrary code leaves room for human errors, and thus increases the chances of bugs \cite{de2017understanding, luu2016making,gatteschi2018blockchain}. In sum, the degree of automation depends on the services provided which is closely linked to the \emph{design} of a blockchain platform but at the cost of additional complexity.

\item \textit{Transparency $\leadsto$ Auditability}. Each node in a blockchain P2P network holds a complete copy of the distributed ledger making all transactions transparent \cite{DoyouneedaBlockchain?2017, howblockchain, swan2015blockchain, gatteschi2018blockchain}. However, for permissioned blockchains permissions to read the ledger can be confined to increase transaction privacy. Decreasing the transparency of the transaction records makes permissioned blockchains less auditable \cite{xu2017taxonomy, deblockchainsforgovernmental2017distributed, DoyouneedaBlockchain?2017}. In short, the auditability of the network depends on the permission \emph{arrangement} of the P2P network.

\item \textit{Immutability $\leadsto$ Verifiability}. The entire history of transactions performed is recorded and stored in blocks. Given that these blocks are cryptographically chained using hashes, the record becomes immutable \cite{realizingthepotentialof2017,arevolutionintrust2017,applicationsofdistributed2017,deblockchainsforgovernmental2017distributed,DistributedLedgerTechnologyandBlockchain2017}. Provided that the entire history of transactions is auditable, the proof that any transaction has (not) taken place in the past is thus verifiable since blockchains are append only \cite{applicationsofdistributed2017,distributedledgertechnologyforthefinancialindustry2016,gatteschi2018blockchain}. An insecure consensus protocol that allows for the introduction of blocks containing double spend transactions could jeopardize the immutability of the ledger. Therefore the immutability of a blockchains distributed ledger depends on how transactions are \emph{processed} during the consensus protocol \cite{arevolutionintrust2017,yaga2018blockchainoverview,DistributedLedgerTechnologyandBlockchain2017}.
\end{itemize}

\section{Blockchain Technology: Challenges and Outlook}
\label{challoutlook}

Despite being a promising novel technology, currently BCT faces several challenges that inhibit widespread adoption. This section highlights and discussses the challenges evident from the literature.

\subsection{Latency} One of the challenges BCT faces is that most consensus protocols have a high latency, meaning that the time between the submission of transactions and their confirmation is high \cite{xu2017taxonomy,globalblockchain2017,gatteschi2018blockchain,meng2018intrusion,glaser2017pervasive, mougayar2016business}. This is due to the fixed blocktime interval for most blockchain networks. Effectively this means that on average it takes the Bitcoin network roughly 60 minutes before transactions are settled and can be regarded as final \cite{globalblockchain2017, xu2017taxonomy}. Ethereum has made significant process in this area using the \emph{GHOST} (Greedy Heaviest Observed Subtree) protocol by increasing the block interval to 14 seconds and transaction finality after 12 blocks \cite{xu2017taxonomy,gatteschi2018blockchain}.

What is more, currently  the finality for clearing and settling transactions is a legally defined moment. When using BCT to enact transactions settlement finality is probabilistic; The longer a transactions is considered settled by network participants, the less likely it will become that the transaction will be reversed or declared invalid \cite{distributedledgertechnologyinpayments2016,globalblockchain2017}. Clearly, these two arrangements are at odds. A direction that is currently being explored to concurrently address the throughput and latency issues of BCT is that of \emph{sharding} the mining network. ELASTICO \cite{luu2016sharding} is an example of a consensus protocol that shards the mining network. When sharding the network miners are uniformly partitioned into smaller committees that process a specific set of transactions. Accordingly transactions can be processed in parallel and thus throughput capacity can be increased. 

\subsection{Throughput} The maximum throughput of transaction has also been shown to be a challenge \cite{xu2017taxonomy,DoyouneedaBlockchain?2017,globalblockchain2017,mingxiao2017review,Zheng2017AnOverviewofBlockchainTechnologyArchitectureConsensusandFuturetrends}. The concurrent throughput challenges for BCT are closely related to those of the latency. At the time of writing the Bitcoin network can reach a throughput of 7 transactions per second \cite{globalblockchain2017,dinh2018untangling}.

Yet again this problem is related to the blocktime interval but also to \emph{blocksize}. The size of a block determines how many transactions can be included. For the Bitcoin the size limit of a block is 1 MB \cite{xu2017taxonomy}. Recently there have been proposals to increase the throughput of the Bitcoin blockchain by increasing the blocksize from 1MB to 8 MB \cite{xu2017taxonomy}. Proponents and opponents of this proposal have interchanged various arguments that so far has reached no conclusive upper-hand \cite{realizingthepotentialof2017}. By implementing the \emph{GHOST protocol} Ethereum has managed to improve it's throughput capacity to 15 transactions per second because the block time interval is smaller (14 seconds). Rather than following the longest chain, in GHOST a miner weights the branches in terms of the computational power spend to create them and chooses the better one to follow. Another promising novel development is the introduction of \emph{off-chain payment channels} such as Raiden\footnote{www.raiden.network}, Bitcoin Lightning\footnote{www.lightning.network} and Sprites \cite{miller2017sprites} that enables two parties to directly and privately maintain a two-party micro payment channel. Khalil and Gervais \cite{khalil2017revive} extend the concept of off-chain payment channels by suggesting a novel approach that enables the refunding of existing payment channels when they are depleted without performing a transaction on the blockchain network. Recently the segregated witness (SegWit) proposal has been suggested in the Bitcoin community to change the internal design of blocks to increase the throughput of transactions. The proposal entails separating (segregate) signatures (witnesses) from the remainder transaction data. In this manner the size of the witnesses does not add to the data size limit of the blocks \cite{xu2017taxonomy}.

\subsection{Data Storage} Another challenge that is pointed out by both practitioners and scholars alike is how to cope with the evergrowing need for \emph{data storage space} \cite{globalblockchain2017,swan2015blockchain,gatteschi2018blockchain, lee2018bidaas,zohar2015bitcoin}. This challenge mainly stems from the fact that to verify transactions, a node needs to be aware of the whole blockchains' history. If the Bitcoin were to process an equal number of transactions as Visa the amount of storage required would grow by 214 PB per year \cite{swan2015blockchain}. Some suggestions for improvement of data storage have been made such as the introduction of \emph{lightweight clients} that do not download the complete record of transactions. Instead, lightweight clients download only the blockheaders to validate transactions. To verify transactions these nodes use a technique called Simplified Payment Verification (SPV) \cite{yaga2018blockchainoverview,Zheng2017AnOverviewofBlockchainTechnologyArchitectureConsensusandFuturetrends}.

\subsection{Data Privacy} \label{datapriv}Preserving privacy of participants and confidentially of their data has turned out to be a fundamental challenge \cite{BlockChainTechnologyBeyondBitcoin2015,Zheng2017AnOverviewofBlockchainTechnologyArchitectureConsensusandFuturetrends,xu2017taxonomy,mendling2018blockchains,applicationsofdistributed2017,DoyouneedaBlockchain?2017,globalblockchain2017, mougayar2016business}. Although transparency is one of the key characteristics of especially public blockchain networks it is at odds with privacy. For public blockchains by design every transaction needs to be visible to every participant for the sake of public verifiability \cite{globalblockchain2017,androulaki2018hyperledger,yaga2018blockchainoverview}, though they can be encrypted and the identity of the user is hidden. In order to address this problem private and consortium blockchains such as Hyperledger \cite{androulaki2018hyperledger} and Corda \cite{brown2016corda} with a permissioned model have been introduced \cite{DistributedLedgerTechnologyandBlockchain2017}. Another approach solve this problem is the usage of mixers and ZKP (see \ref{zkp}) \cite{bailis2017research}. In a study \cite{globalblockchain2017} 57\% of the respondents stated that implementing privacy-enhancing techniques in their BCT systems is planned for the future. Out of these respondents 78\% have expressed the desire to implement zero-knowledge proofs (ZKP). A second privacy challenge is that a blockchain ledger is \emph{immutable}; Once a transaction has been stored in a block it can not be removed. Further, in permissionless blockchain every node is able to view all transactions and, consequently explore the entire history of transactions. The General Data Protection Regulation (GDPR) however, enforces restrictions on how information about EU citizens may be used and stored \cite{olnes2017blockchain}. One of the rules that would be difficult to comply with is the "right to be forgotten" that allows an individual to demand the erasure of information under certain conditions. Clearly, the immutability of a blockchains ledger is incongruent with the right to be forgotten \cite{DistributedLedgerTechnologyCybersecurity2016,applicationsofdistributed2017}.             

\subsection{Governance}
\label{challenges_governance}
The governance of a blockchain with regards to \emph{updating} its fundamental rules is problematic \cite{yu2018virtualization, yaga2018blockchainoverview,DistributedLedgerTechnologyandBlockchain2017,zohar2015bitcoin}. A prime example is the ongoing debate within the Bitcoin community about the block size which has ended in a stalemate \cite{realizingthepotentialof2017, deblockchainsforgovernmental2017distributed}. Even for centralized systems updating software can be difficult let alone when a system has many users, geographically dispersed, as can be the case with BCT \cite{yaga2018blockchainoverview}. Another classic example of the governance problems blockchain currently faces is the response of the Ethereum community to the DAO hack\footnote{The term hack must be qualified; The attacker exploited a vulnerability in the smart contract that allowed a split function (enabling the withdrawal of funds from the contract) to be called repeatedly in order to withdrawal more funds than entitled to. For further reading about the DAO hack the author recommend reading Annex B in \cite{DistributedLedgerTechnologyCybersecurity2016}.}. Due to unintended flaws in the semantics of a contract an attacker was able to obfuscate a large amount of Ether worth an estimated \$50 million\footnote{Tapscott and Tapscott \cite{realizingthepotentialof2017} argue that the total worth of the obfuscated Ether was around \$70 million, whereas Gatteschi et al. \cite{gatteschi2018blockchain} suggest that it was the equivalent of \$60 million.}. In response to the attack, a hard fork was proposed to recover the Ether, to which 89\% of Ether-holding voters gave their consent. Some of the remainder non-consenting voters rejected this fork of the blockchain mostly for philosophical reasons, including the principle that a blockchain is immutable. These voters decided to use the unforked Ethereum blockchain resulting in a split into two separate currencies: Ether (containing the hard fork) and Ethereum Classic (no hard fork) \cite{yaga2018blockchainoverview,DistributedLedgerTechnologyandBlockchain2017}. Another governance issue that needs to be addressed is that of \emph{key management}; BCT is decentralized and as such when a user forgets their private key there is no central authority to recover it \cite{DistributedLedgerTechnologyCybersecurity2016}. As a solution to this problem He et al. \cite{he2018social} present a wallet-management system based on semi-trusted social networks to recover wallets and the keys they hold. However, the true Achilles heel with regard to private key management is related to the wallets that store these key's; If the hardware on which a users wallet get's lost, targetted with malware or is attacked the private key might get lost or stolen \cite{bailis2017research}.

\subsection{Usability} A more practical challenge that hampers the widespread adoption of BCT is the current lack of end-user support (BCT is hard to use and to understand) and adequate developer support (few developers tools available) \cite{mendling2018blockchains, mougayar2016business}. In line with these observations, research by Tapscott and Tapscott \cite{realizingthepotentialof2017} indicates that many Dapps are not accessible to the average person and that interfaces are user-unfriendly. Further, in their study they suggest that there are approximately between 1000 and 2000 developers that understand how to develop Dapps. However, one of their interviewees stated that this number could perhaps increase by establishing creative educational programmes.  

\section{Discussion}
\label{discussion} 

Our GT-based analysis was used to populate the illustrated 4+1 views, properties and challenges of BCT. 
First, this section grounds the insights on the 4+1 views of BCT architecture elements, properties, and challenges through discussion. More specifically, the GT-driven 4+1 perspectives on BCT are deepened by discussing coding frequencies and trends of the concepts in literature. 
Secondly, the section presents observations we made in the scope of our analysis derived from examining the distribution of the topics found in the sample and synthesizing their contents.  

\subsection{A Grounded-Theory of Blockchain Technology}
\label{sec_concepts}

\begin{figure}
    \centering
    \includegraphics[width=\textwidth]{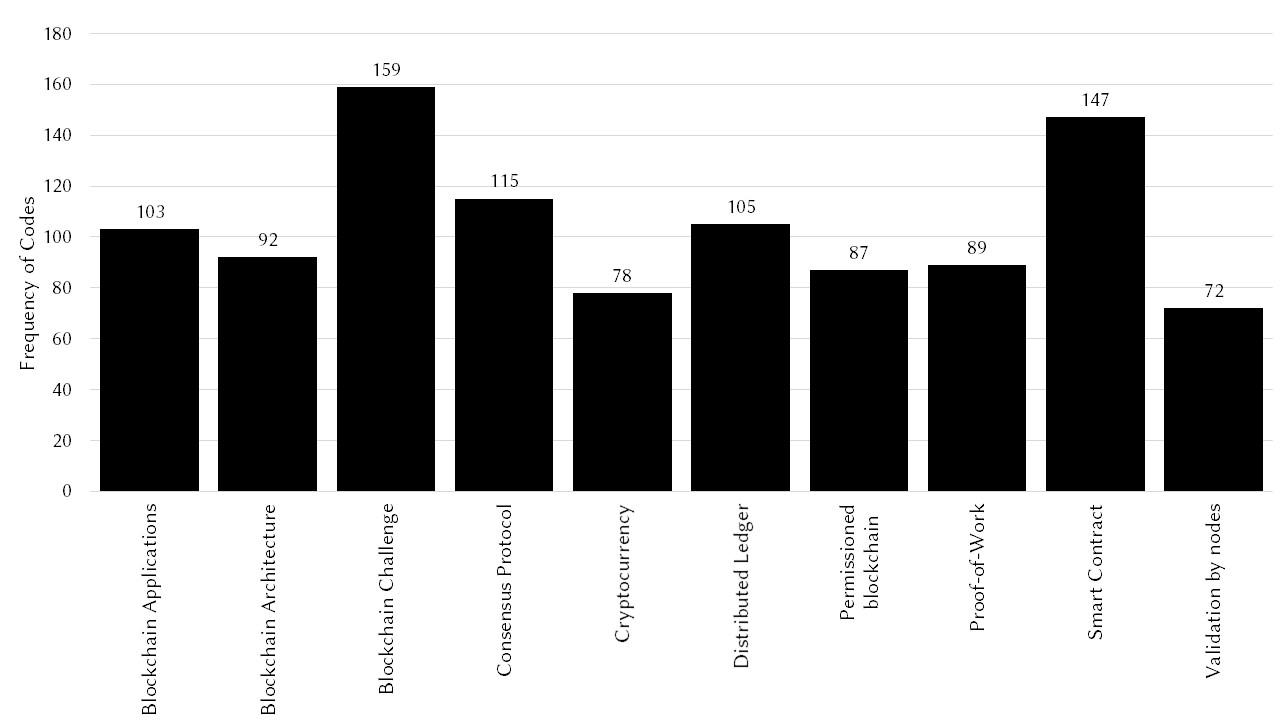}
    \caption{Frequency of top 10 most recurring codes.}
    \label{stats_codes}
\end{figure}

The first step of in the data analysis procedure was to apply an open code each time the literature reflected a concept (see Sec. \ref{Data Analysis}). A frequency analysis of the open codes (how often certain codes have been applied) unravels which concept are deemed important. Figure \ref{stats_codes} depicts the top 10 most frequently used codes.

Blockchain challenges were mentioned most, followed by smart contracts, and consensus protocols. These results show that blockchain challenges are widely discussed in the papers under review and that the technology has not yet come to full fruition. The frequency analysis further showed that smart contracts are most frequently discussed, and can therefore be deemed a key concept for BCT future evolutions.

The third most often applied code is that of consensus protocols, and among them PoW. These findings resonate with the number of papers on the topic of PoW consensus protocols (see Sec. \ref{sec: results}). Taken together, these findings show that literature is still predominantly focused on the PoW consensus protocol, whereas several other consensus protocols nowadays exist. 

Surprisingly, the term permissioned blockchain is more often mentioned than permissionless blockchain which is not mentioned as one of the top ten concepts. However, these results can also be attributed to the perception that blockchains generally are public and permissionless and as such that permissioned blockchains are an exemption to be specifically mentioned.

\subsubsection{Grounding the Logical View of BCT}
\label{gtlogicalviewbct}

\begin{figure}
    \centering
    \includegraphics[width=\textwidth]{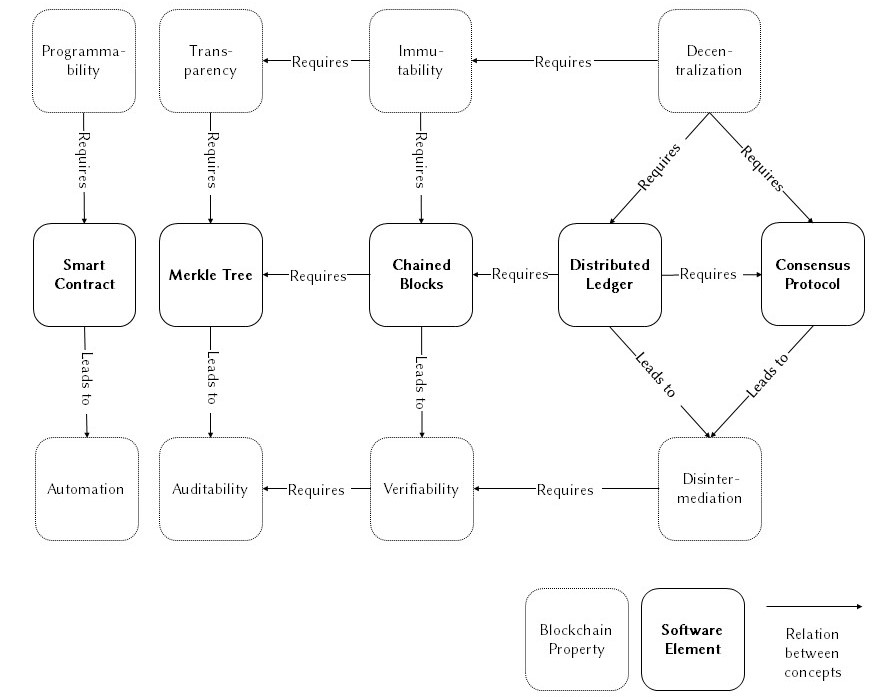}
    \caption{BCT software elements and properties, an overview.}
    \label{fig_elementsprop}
\end{figure}

BCT encompasses several software elements that combined create the architectural properties of BCT. The results of the axial coding revealed what the intricate relation among software elements and properties (see Fig. \ref{fig_elementsprop}).

The software elements of BCT work in concert to allow for secure transactions on a P2P network (see Sec. \ref{logicalview}). We found that there is a strong dependency among these elements which has been depicted in Fig. \ref{fig_elementsprop} by the middle row of blocks: During a consensus protocol nodes verify transactions, which is not possible without the availability of a distributed ledger. In turn, the distributed ledger employs the concept of chained blocks that depends on a Merkle Tree to summarizes the transactions. Because each software element is implemented to yield a particular property (see Sec. \ref{charactbct}) these properties also are also connected. This relation is shown as the top row of blocks in Fig. \ref{fig_elementsprop}. 

Decentralization requires a consensus protocol and distributed ledger but in addition immutability of the transaction records. The immutability of transaction records is dependent on the degree of transparency of a ledger (see Sec. \ref{charactbct}). However, transparency does not only guaranty the immutability of the distributed ledger, but also leads to auditability of the transaction records (bottom row Fig. \ref{fig_elementsprop}). Verifiability of a blockchain requires auditability as without no proof can be provided that a transaction has not already been spend. 

An interesting observation is that programmability does not seem to have a direct relation to the other properties of blockchain. Every blockchain has a certain degree of programmability \cite{xu2017taxonomy}. However, arbitrary programmability of transactions requires the concept of smart contracts which is optional for blockchain design \cite{xu2016connectblockchain}. The discussion of these results illustrates the complex relation among BCT software elements and properties. Considering these relations will be of importance to obtain the appropriate design when implementing BCT.

\subsubsection{Grounding the Process View of BCT}

\label{gt:processviewbct}

\begin{figure}
    \centering
    \includegraphics[width=\textwidth]{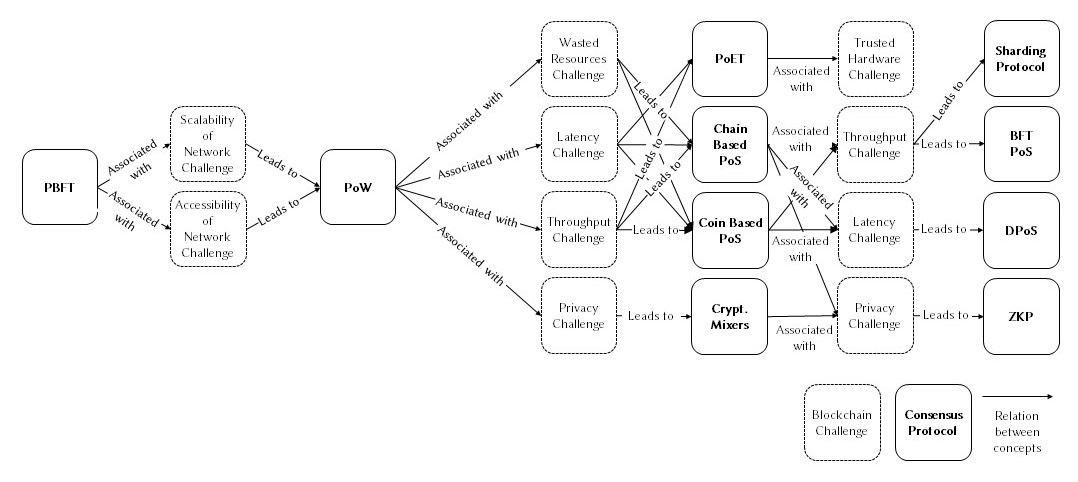}
    \caption{Consensus protocols related to challenges, in chronological order from left to right.}
    \label{fig_consensuschallenges}
\end{figure}

Another relation that emerged from the axial coding process is the relation between BCT challenges and consensus protocols. In Fig. \ref{fig_consensuschallenges} this relation is shown simultaneously with the introduction of consensus protocols in chronological order.

Our results from the axial coding process show that each consensus protocol developed after the PoW protocol aims to tackle a particular BCT challenge. Furthermore, unsatisfied with prior protocols to address the challenges of BCT novel consensus protocols were introduced over time. For instance, the development of cryptographic mixers clearly aimed at providing more transaction privacy. To facilitate fully private transactions however, ZKP based protocols were introduced. While both approaches increase or facilitate fully private transactions these protocols were not designed to achieve faster throughput as compared to PoW.

On the other hand, PoS based protocols have aimed at improving throughput and latency performance as compared to PoW. Among the first of these attempts were the coin and-chain based PoS variants (see Sec. \ref{PoS}). As the performance aspects of these PoS variants were still considered unsatisfactory BFT based PoS, DPoS and sharding based protocols were introduced \cite{li2017securing, yaga2018blockchainoverview}. Neither the PoS, DPoS or sharding consensus protocols address privacy of transactions.  

Taken together the results show that there are have been desperate attempts to address BCT challenges. A uniform approach that simultaneously addresses specific sets of problems is still lacking. This implication might has some important ramifications for future research efforts, for example, several research efforts by practitioners and scholars have led to consensus protocols that only pursue to improve one favorable property while a unified agenda to develop a consensus protocol that is designed with both privacy and performance aspects (throughput and latency) in mind is still lacking. 

\subsubsection{Grounding the Physical View of BCT}

\begin{figure}
    \centering
    \includegraphics[width=\textwidth]{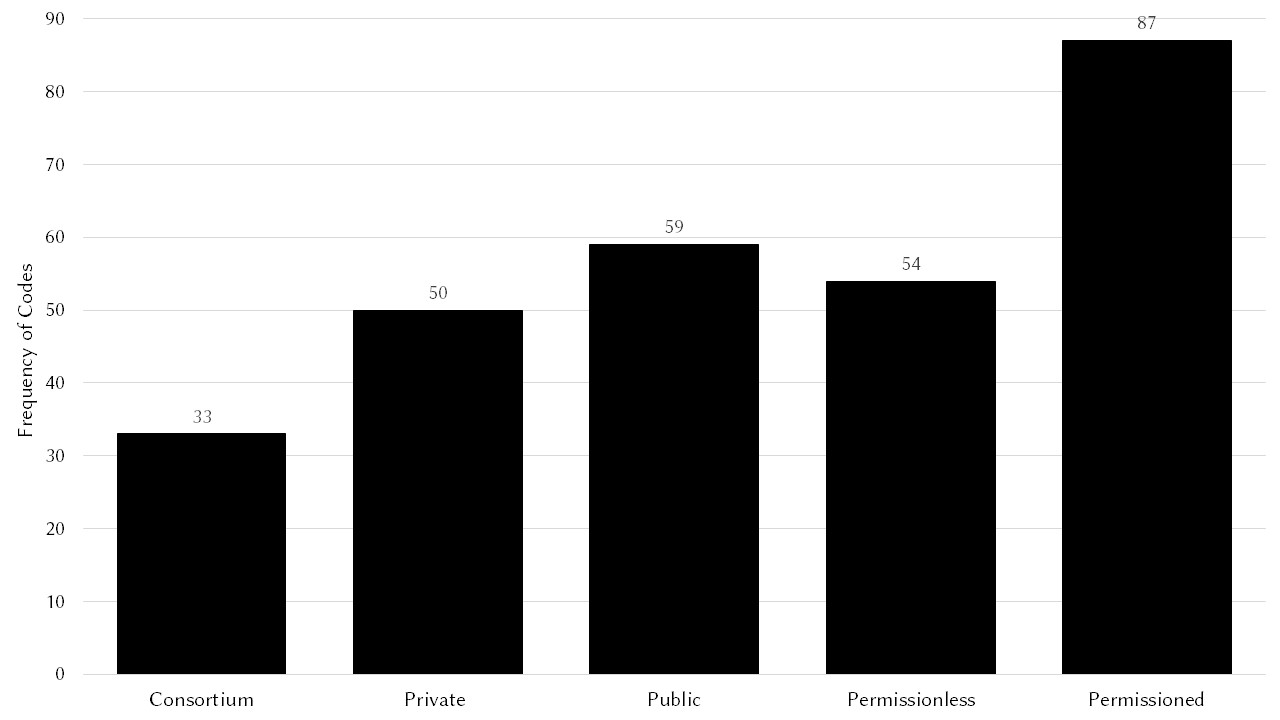}
    \caption{BCT network types, coding frequency.}
    \label{stats_bctnetworks}
\end{figure}

The grounded view describes the design patterns of a blockchain network. As discussed in Sec. \ref{processview} the design of the P2P network that supports the blockchain can differ. The frequency of the codes concerning BCT networks have been examined (see Fig. \ref{stats_bctnetworks}) to establish a notion of their importance.

What becomes clear from these statistics is that surprisingly permissioned blockchain networks are more mentioned than permissionless BCT networks. A closer examination of the papers in which the codes were used reveals that permissioned networks are predominantly employed to contrast the properties (negative and positive) of permissionless networks. Public networks are the third code that is most often applied, followed by private networks. These paradoxical networks are discussed in tandem to contrast their properties while Consortium networks are mentioned least. It must be noted however, that the terms consortium and private networks are often discussed in the same context.

\begin{figure}
    \centering
    \includegraphics[width=\textwidth]{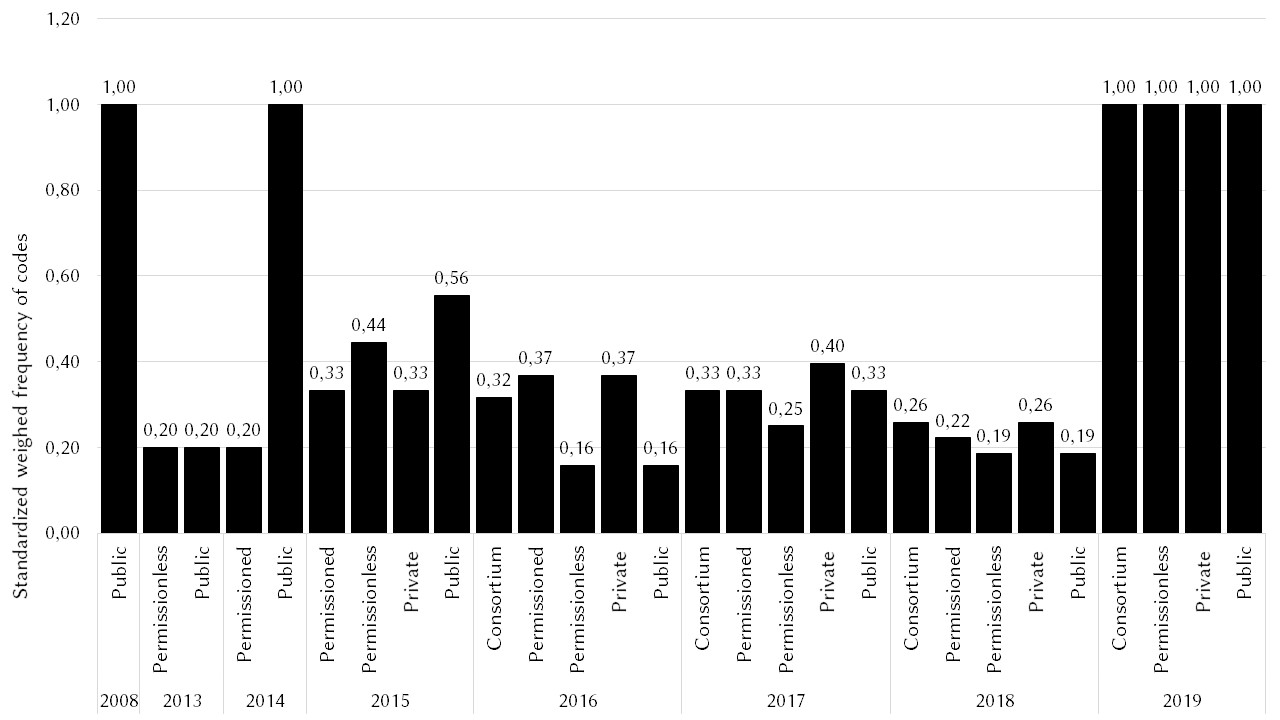}
    \caption{Trends in BCT networks from 2008 to 2019.}
    \label{fig_stats_bctnetworks_trends}
\end{figure}

The trends on BCT networks also sheds an important light on the developments in the field of BCT. These trends are shown in Fig. \ref{fig_stats_bctnetworks_trends} and have been constructed by counting each time a paper was coded the network type in a given year. Thereafter the result has been divided by the total number of papers in the sample for a particular year. We deemed this last step appropriate to account for the fact that over time more work on BCT has been published.

Analysis of the trends regarding BCT networks codes (see Fig. \ref{fig_stats_bctnetworks_trends}) showed that in time more papers were mentioning private and consortium networks combined with permissioned models. When closer examining this trend we found that over time more papers were discussing private, consortium and permissioned networks. The literature under review \cite{pilkington2016, arevolutionintrust2017, gatteschi2018blockchain, dinh2017blockbench, globalblockchain2017} states that these network types were introduced as a response to the current challenges public oriented networks face (e.g. privacy and throughput). Therefore this trend can be explained by the fact that in time public blockchain challenges and limitations became more evident and thus alternatives were introduced.

\subsubsection{Grounding the Use-Case View of BCT}
\label{gt:usecaseviewbct}
The axial coding process of papers concerning BCT applications revealed that these applications have a distinct focus to which they have been categorized (see Fig. \ref{fig_maintopics}). A large strand of literature is focused on cryptocurrencies and improving the interoperability between chains using different coins, or fluctuations in the prices of cryptocurrency. The focus of these papers lies on \emph{transactions} carried out using a blockchain or between chains and what influences these transactions. For these papers therefore the cryptocurrency itself becomes the \textit{specific} focus of study as they are the embodiment of blockchain transactions. This contrast with other papers that see cryptocurrency as a means to an end and tend to have a more general focus on employing the technology as a whole. Several papers suggest the use of BCT to enhance the utility of IoT devices. Here, BCT \emph{primary} serves to improve, amongst things, security and connections of these devices. In turn, IoT devices could potentially be used for a wide array of other applications. Discerning whether BCT \textit{is} the application or rather a means to an end helps to categorize papers in the field of BCT.

Another observation resulting from the axial coding process is that the general application categories found in the literature sample utilize different properties of BCT. Record keeping applications mostly benefit from the decentralized nature of BCT. Whereas applications related to security and fraud detection seem to employ BCT for the sake of verifiability. For securities and insurance applications the auditability of the transactions is most important. Thus, besides categorizing applications based on their primary focus, one should also taken into account which of BCT's properties are predominantly utilized. This observation is important to take note of when developing BCT based applications as some properties can be deemed more valuable for the design than others and can therefore traded-off against each other. 

\subsubsection{Grounding the Properties and Challenges View of BCT}
\label{gt:propertiesandchallenges}

\begin{figure}
    \centering
    \includegraphics[width=\textwidth]{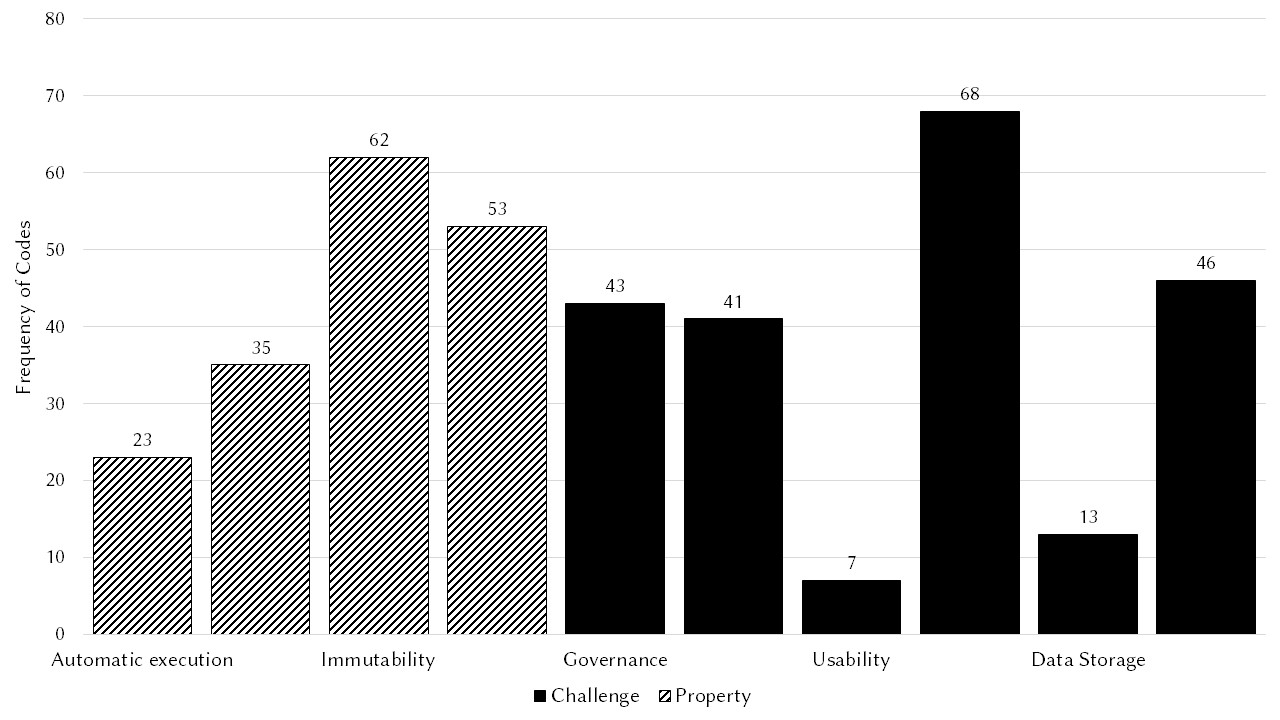}
    \caption{Coding frequency of BCT challenges and properties.}
    \label{stats_characprop}
\end{figure}

BCT has been developed to attain certain properties yet also gave rise several challenges. We juxtaposed the results of a frequency analysis conducted using the codes applied for BCT properties and challenges to discuss their relation. The frequency of the codes related to BCT properties and challenges can be found in Fig. \ref{stats_characprop}.

Challenges related to privacy on a blockchain is mentioned most of all codes related to challenges. The immutability is the most frequent applied code related to BCT properties. Paradoxically, whereas privacy the most discussed challenge for BCT, the transparency of the ledger is mentioned as the second most important property of BCT. The decentralized nature of BCT is the single least applied code. Automatic execution is the least recurring code of related to a property of BCT. However, papers that predominantly discuss properties from the perspective of the Bitcoin do not mention smart contracts which is related to the concept of automatic execution.

A further analysis of the relation between the throughput and latency codes applied reveals that the coding of these two concepts coincides, resulting in an almost equal number of times these codes are applied. Governance of blockchain networks as a challenge is almost as frequently discussed as the technical issues of BCT (throughput and latency). The data storage and usability codes are less frequently used in comparison to the other codes concerning BCT challenges. Of these two codes usability is least often mentioned as a challenge. These results can be explained by the fact that most papers either focus on the applications for BCT or on architectural aspects without regarding the users perspective.

\begin{figure}
    \centering
    \includegraphics[width=\textwidth]{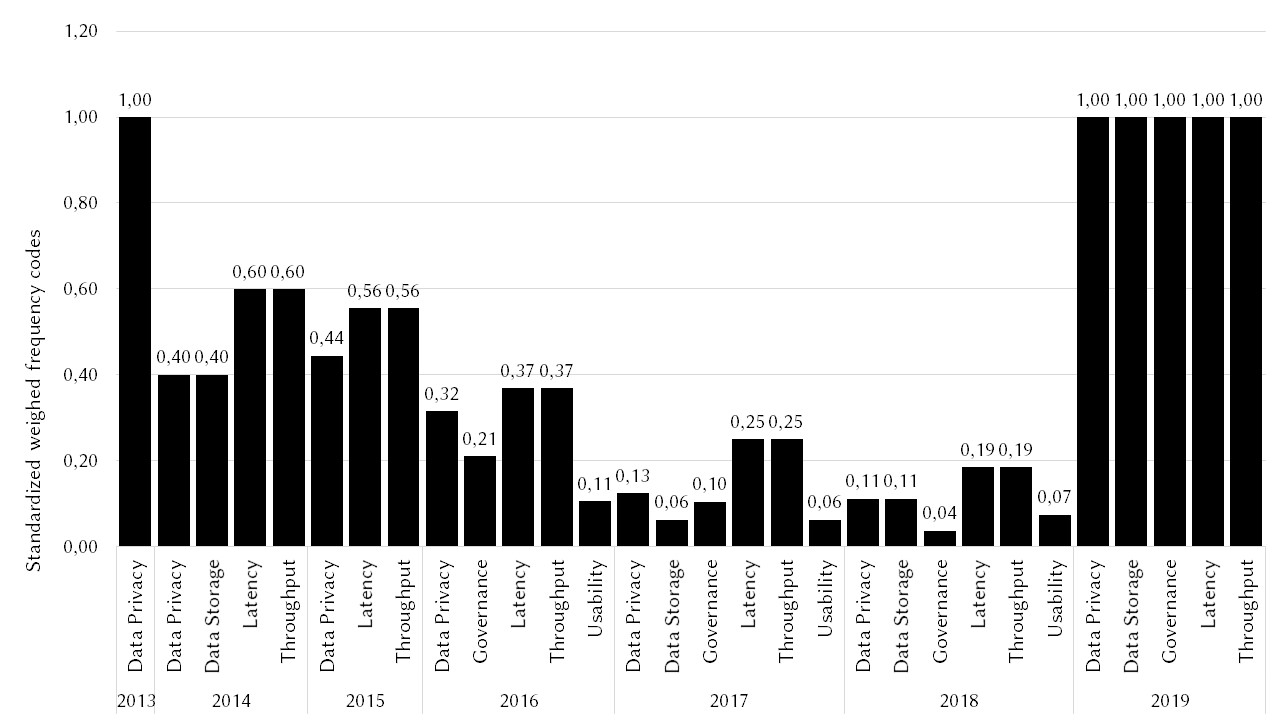}
    \caption{Trends in BCT challenges from 2008 to 2019.}
    \label{fig_trends_challenges}
\end{figure}

An analysis of the applied frequency of codes over time reveals the trends in BCT challenges. Again, the analysis has been conducted by counting each time a paper was coded a certain challenge in a given year. After obtaining the results these have been divided by the total number of papers for a particular year. The weighted number of codes per are challenge and year are depicted in Fig. \ref{fig_trends_challenges} to provide a fine-grained perspective. 

The figure shows an overall decline between 2008-2019 in the mentioning of all challenges\footnote{For the years 2013 and 2019 only one paper has been included in the literature sample which imbalances the standardized weighted frequency of the codes for these years. Hence, in the scope of this analysis the results for 2013 and 2019 should be regarded as outliers.}. What is interesting however, is that when the codes are weighted (I.e. per year in relation to total number of papers) the results demonstrate that throughput and latency are the most important challenges to address. From publications in 2016 challenges related to BCT governance and usability first emerge. This shows that from 2016 onwards blockchain became mainstream adopted as the usability of BCT (for non-programmers) for the first time were discussed in \cite{walport2016distributed,distributedledgertechnologyinpayments2016}. Moreover, in that same year governance emerged from the literature under review \cite{embracingdisruption2016, distributedledgertechnologyinpayments2016, peters20166understanding, walport2016distributed} as a challenge. This literature discusses that novel updates to existing platforms were required due to several reasons yet that this turned out to be a challenge. In tandem, the governance issues from a legal perspective are discussed which means that BCT is was no longer regarded as a novelty but mainstream and a technology that needed to adhere to legal standards (see Sec. \ref{challenges_governance}).

\subsection{Highlights and Observations}
\label{disc:highlightsandobservations}
\subsubsection{Chaining data using blocks}

From the FCA conducted for this research we found that what sets blockchain apart from other forms of distributed ledgers is that transactions are stored in blocks. On the one hand, transaction data is stored in blocks to ensure the integrity of the distributed ledger. On the other hand, storing data in blocks also has its drawbacks. 

Blocks have fixed data sizes and can only contain a limited number of transactions. Transaction that do not fit in the current block that is created have to wait to be processed and included in the next block. Given that both blocktime and blocksize for most blockchain networks are fixed these two parameters determine throughput capacity of a blockchain network \cite{DistributedLedgerTechnologyandBlockchain2017}. The SegWit proposal (see Sec. \ref{challoutlook}) might partially solve this problem by expanding the size of blocks. However, to increase the throughput of public blockchain protocols further future expansions of the blocksize will be required. What the consequences are for the security of the network remains unknown however, because there has been no empirical investigation to test these configurations. 

Nodes on a blockchain network need to keep a complete history of all transactions made on a blockchain network in order to validate them. Because the number of transactions that have been made on a blockchain is growing over time the data storage demands grow in parallel. Consequently, on the long-term it will be unsustainable for every node to keep the entire history of transactions \cite{swan2015blockchain}. Conversely, if only few nodes would be able to meet the store demands it would defeat the purpose of decentralization. A similar trend can be observed with regards to Bitcoin mining. Initially nodes mined blocks individually. However, due to increased mining requirements (CPU power) they eventually started collaborating in mining pools and mining became more centralized as a result. To address this problem some blockchains are utilizing the concept of checkpoints \cite{li2017securing, deblockchainsforgovernmental2017distributed}. Whether this solution aides in securing the network from attacks remains unknown. Another proposed solution is the use of lightweight clients. However, at least some nodes need to be full weight clients that keep track of all transactions thus, still have to burden themselves with storing large amounts of data. 

\subsubsection{Consensus in the wild}

Besides the consensus protocols discussed in the results section (see Sec. \ref{sec: results}) we reported more consensus protocols such as \emph{Proof of Work or Knowledge} (PoWorKs) \cite{baldimtsi2016indistinguishable}, \emph{Proof of Vote} (PoV) \cite{li2017proof}, \emph{Proof of Sequential Work} (PoSW) \cite{cohen2018simple}. However, none of these protocols is supported by mature implementations, practical application, or empirical evidence of operational characteristics. For the sake of completeness, however, Tab. \ref{tabbaoverview} offers an overview of all primary consensus protocols (without any derivation, e.g., PoSW with respect to PoW) and their claimed features in the respective literature. The table articulates every consensus protocol (ID in column 1) using (a) the trust-level required in the P2P network setting, (b) the scalability (i.e., whether the network can scale in terms of the number of nodes \cite{Kalinov06}), (c) the throughput, i.e., how many transactions per second the protocol can successfully process and (d) the latency, i.e., the time it takes to successfully confirm the transaction.

\begin{table}

\footnotesize
\caption{A complete reference over consensus protocols; instances are described along their characteristics, an implementation example and the source for argument of the claims.}\label{tabbaoverview}
\begin{tabular}{|>{\centering}p{.6cm}|>{\centering}p{1cm}|>{\centering}p{1.2cm}|>{\centering}p{2cm}|>{\centering}p{2cm}|>{\centering}p{1cm}|>{\centering}p{1.5cm}|>{\centering}p{1.5cm}|}
\hline 
\textbf{ID} & \textbf{Trust-level} & \textbf{Scalability} (\#Nodes) & \textbf{Byzantine Fault-Tolerance} & \textbf{Throughput} & \textbf{Latency} & \textbf{Example} & \textbf{Source}\tabularnewline
\hline 
PBFT & High  & Weak  & < 33.3\% of faulty replicas  & <2000  & <10s & Hyperledger Fabric v0.6\footnote{www.hyperledger.org/projects/fabric}
& \cite{Zheng2017AnOverviewofBlockchainTechnologyArchitectureConsensusandFuturetrends,mingxiao2017review,cachin2017blockchains}\tabularnewline
\hline 
RAFT & High & Weak & <51\% of faulty nodes & >10k & <10s & Corda\footnote{www.corda.net}
& \cite{mingxiao2017review,cachin2017blockchains}\tabularnewline
\hline 
PoW & Low & Strong & <51\% of computing power & <100 & >100s & Bitcoin\footnote{www.bitcoin.com} 
& \cite{mingxiao2017review,Zheng2017AnOverviewofBlockchainTechnologyArchitectureConsensusandFuturetrends}\tabularnewline
\hline 
PoET & Low & - & <51\% of computing power  & - & - & Hyperledger Sawtooth\footnote{https://www.hyperledger.org/projects/sawtooth} 
& \cite{cachin2017blockchains}\tabularnewline
\hline 
PoS & Low & Strong & <33\% of stake & <1000 & <100s & Tendermint\footnote{www.tendermint.com} 
& \cite{Zheng2017AnOverviewofBlockchainTechnologyArchitectureConsensusandFuturetrends,mingxiao2017review,cachin2017blockchains}\tabularnewline
\hline 
DPoS & Medium & Strong & < 51\% of validators & <1000 & <100s & Bitshares\footnote{www.bitshares.org} 
& \cite{Zheng2017AnOverviewofBlockchainTechnologyArchitectureConsensusandFuturetrends, mingxiao2017review}\tabularnewline
\hline 
ZKP\footnote{ZCash uses an Equihash variant of the PoW protocol; for Equihash the Blocktime differs from that of the Bitcoins PoW no direct data is provided in literature on its throughput and latency.}  & Low & Strong  & <51\% of computing power  & - & - & Zcash\footnote{www.z.cash} 
& \cite{dinh2018untangling}\tabularnewline
\hline 
\end{tabular}
\end{table}

An interesting observation with respect to consensus protocols in Tab. \ref{tabbaoverview} is that there seems to be a lack of systematic and empirical studies to test the claims around the proposed consensus protocols and their architectural properties. Of the papers reviewed, none based their statements about these properties and how they behave under different circumstance on the results of empirical testing. An exception is a study by Dihn et al. \cite{dinh2018untangling} that provides a framework to benchmark private blockchains. Public blockchains however, remain untested. Nevertheless some general observations can be made: With the introduction of PoW a shift is made from protocols that require high trust among nodes to a more trustless setting. This is not surprising given that one of the main aims of the original Bitcoin protocol was to make it expandable beyond a fixed number of authenticated nodes. Accordingly, an increase in byzantine fault tolerance of consensus protocols can also be observed. However, the results also demonstrate that an increase in the byzantine fault tolerance is coupled with a decrease in throughput and latency. For instance, the Tendermint PoS variant has a higher throughput and lower latency as the Bitcoins PoW protocol, yet it is also less byzantine fault tolerant (<33\% to <51\%).  

For the PoET and ZKP proof based consensus protocols there is no data available with regard to their properties. Moreover, although it is stated in \cite{dinh2018untangling, internetofthingschristidis2016blockchains} ZK-SNARK techniques incur large overheads in terms of storage space, there is little evidence to substantiate these claims.  

The gradual shift from PoW towards PoS based consensus can have some important ramifications for the security of blockchains. With the exception of \cite{li2017securing} little efforts have been made to investigate the unresolved security issues of PoS variants and how to resolve address them. One such problem is how to determine the deposit forgers are required to make to participate in the consensus protocol (see Sec. \ref{PoS}). If the gains of introducing invalid transactions outweigh the losses of a deposit the solution is will not be effective. 
Sharding of consensus seeking is a novel development mentioned in the papers under review. Yet we view that there are many unaddressed questions concerning blockchain sharding. When sharding a consensus protocol the nodes in the network will be distributed between several shards. How this distribution should take place remains unclear however. Provided that a secure manner has been found to distribute the nodes then it should be determined how many nodes must encompass a shard to make it secure as small shards (with few nodes) are easy to attack. 

\subsubsection{Blockchain Hybrids Emerging}

Our literature suggests that more recently several interesting combinations of public blockchains with a permissioned model are under experimentation. 

We observe from the coding process that the terms permissioned/private blockchain and permissionless/public blockchain are used interchangeably. Based on our extensive review, we propose that blockchain-oriented networks should be categorized based on both (1) network authentication coupled to permissions (permissioned/permissionless) and (2) accessibility of the network (public, consortium, private). Although the first blockchain network (Bitcoin) was public and permissionless, from the coding process it can be observed that in the papers under review permissioned and private/consortium blockchain networks are often mentioned (see Fig. \ref{stats_bctnetworks}). One of the main reasons remarked in literature is that public blockchains have to utilize consensus protocols that can be regarded as slower than the ones that could be employed for private/consortium blockchains. Furthermore, having more control over the permissions each participant in the network has is another reason mentioned. We observe that private/consortium and permissioned networks are often regarded as a substitute to their public counterpart because of these challenges. 

\subsubsection{Mainstream Adoption of Blockchain Technology}
An increase of academic works on the topic, and the overall increase of literature produced every year can be observed (see Sec. \ref{sec: results}). These results show that BCT has been embraced by the mainstream. The usability of BCT for end-users and developers \cite{mendling2018blockchains} could however, hamper the further adoption of BCT. Tools to develop blockchains/smart contracts seem to be absent. Especially for users that aim to employ smart contracts to define simple transaction logic a more simple and user friendly approach that does not involve programming would benefit the mainstream adoption of smart contracts and blockchain.

The results of this research show that governance of blockchain both in terms of the general ecosystem and at the platform level are a challenge currently. In many countries the use of BCT for applications or cryptocurrency has an opaque legal status. Making policy and regulation is difficult since a blockchain networks operate internationally and are not bound to a single jurisdiction \cite{realizingthepotentialof2017, mougayar2016business}. Implementing updates or changes for a public blockchain network are a platform level governance challenge caused by decentralized decision making involving many participants. A solution to this problem might be the centralization of update permissions. However, that solution would be at odds with decentralized principles of blockchains and introduce several security hazards.   

\subsubsection{Safe Executability of Legally-binding Smart Contracts}
Smart contracts are an important application for BCT as can be inferred from the number of times the code has been applied (see Sec. \ref{sec_concepts}). All primary studies focus on discussing the security of Ethereum smart contracts. This trend is highlighted also in related work \cite{BlockchainbMapping}. The most obvious implication of this shortcoming is that more rigorous, generalisable, and formalized approach to analyzing the safe and secure executability of smart contracts is required. A preliminary investigation of safe concurrent smart-contracts' executability Dickerson et al. \cite{dickerson2017adding} who describe an abstract solution to this problem without any rigorous evaluation. The trend towards sharding blockchain consensus can have implications for the execution of smart contracts what these exactly are remains unknown till date. Conversely, the word ``contract" in the smart term definition contract would imply for its legally-bound enforceability under specific operational conditions \cite{clack2016smartfoundations}. 

\subsubsection{Blockchain: Properties versus Challenges}

\begin{figure}
    \centering
   \includegraphics[width=\textwidth]{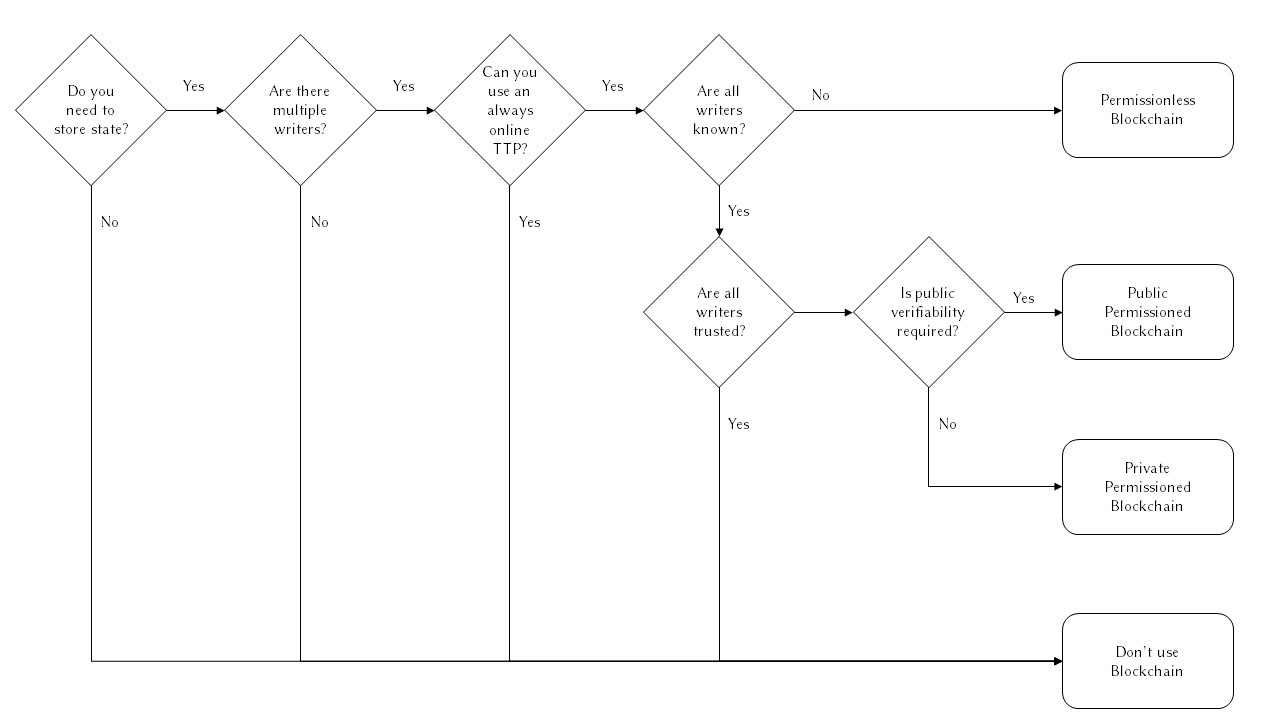}
    \caption{Decision-making model for blockchain networks; adapted from: \cite{DoyouneedaBlockchain?2017}.}
    \label{fig_decisionmodel}
\end{figure}

BCT offers some interesting properties such as disintermediation, programmability, transparency, and verifiability that could potentially be beneficial for many applications (Sec. \ref{charactbct}). However, the technology also has its drawbacks (Sec. \ref{challoutlook}). The work of Mougayar \cite{mougayar2016business} provides a decision-making framework based on business parameters to aid in identifying the problems that BCT can streamline. Our work strives to provide a technical perspective to compound the business side explored by works such as Mougayar \cite{mougayar2016business}. Gatteschi et al \cite{gatteschi2018blockchain} suggest that when considering to adopt BCT one should ask: (1) Whether a shared database is required, (2) Multiple parties write the data, (3) Whether disintermediation is needed (4) and, whether it is required to see the linkage between transactions. In the similar vain W{\"u}st and Gervais \cite{DoyouneedaBlockchain?2017} argue that when deciding whether to adopt and design blockchain the contingencies depicted in Fig. \ref{fig_decisionmodel} should be considered. They argue that if a trusted third party (TTP) always is available there is no need for a blockchain. Thereafter one has to make an analysis of the network participants to decide which type of blockchain network is most appropriate. 

Besides the considerations presented by \cite{gatteschi2018blockchain} and \cite{DoyouneedaBlockchain?2017} from the results of this study we observe the following. First, a blockchain can disintermediate the transactions made on a P2P network between untrusted participants. However, as a result the throughput of most consensus protocols for blockchains is low and their latency high (see. Table \ref{tabbaoverview}). If these properties do not satisfy the requirements of the application these consensus protocols are not the most suitable solution. Visa for instance, handles 2000 transactions per second \cite{swan2015blockchain, distributedledgertechnologyforthefinancialindustry2016} while most protocols currently cannot process transactions at that rate. Furthermore, when all network participants can trust each other a consensus protocol is not required. In such situations other solutions such as distributed databases \cite{OezsuValduriez2011} or P2P network \cite{androutsellis2004survey} offer a less costly and faster way to exchange data or transactions.
Second, not all blockchain platforms allow for the deployment of smart contracts which enable the programmability of transactions. Moreover, currently blockchain based smart contracts present some unresolved problems such as ensuring security \cite{BlockchainbMapping} and legal enforceability \cite{clack2016smartfoundations}. In the past three decades several approaches have been suggested to develop contracts that are executed electronically \cite{lee1988logic, krishna2012methodology} known as electronic contracts that currently have a higher maturity for these applications.
Third, transparency and verifiability are a double-edged sword. In fact, these properties ensure the correctness of the distributed ledger, however transparency is not always desirable. More transaction privacy could be guaranteed by using ZKP's yet it is suggested that these incur additional overhead. An alternative is to perform the transactions on a blockchain network with a permissioned model. This could create censorship resistance which jeopardizes the security of the blockchain (see App. \ref{app_physicalview}).

\subsection{Addressing our Research Questions}

\subsubsection{RQ 1: How can blockchain technology be systematically defined}
The first research question for this study concerned how blockchain technology can be defined --- the RQ is addressed in five consecutive sections. First, in Sec. \ref{bctdef} the study defines BCT based on a FCA. The results of the FCA show that the most mentioned attributes of BCT are transactions, chains, distributed ledger, block, P2P network, consensus protocol, hash function, public key, node and private key. The analysis shows that what sets distributed ledger technology apart from BCT is the storage of transactions in blocks. In presenting this definition we present both scholars and practitioners with an understanding of the BCT phenomenon that can provide common ground for discussion. In Sec. \ref{logicalview} describe and flesh out the interplay between the logical elements of BCT. This description aids in understanding how the combination of individual software elements constitute to BCT. The development view presented in Sec. \ref{developmentview} provides DApp developers and platform developers with some guidance what the main components of BCT are. In Sec. \ref{processview} we elaborate on the workings of consensus protocols. Essentially, our synthesis of the literature shows that increasing the throughput and decreasing the latency of transactions results in a less secure consensus protocol. By providing this overview practitioners can make more informed decisions about the design of their blockchain platform. Furthermore, the section also pointed out the idiosyncratic problems and lack of identification thereof. Scholars can use this information to focus future research endeavors. We are confident that by creating an understanding of current consensus protocols practitioners will be inspired to improve these protocols or, using the lessons learned create new consensus protocols with more favorable properties. The arrangement of the P2P network is discussed in Sec. \ref{physicalview} which has important ramifications for the architectural properties of the blockchain. More informed decision-making can aid practitioners in making the optimal choice when adopting blockchain technology. 

\subsubsection{RQ 2: What applications of blockchain technology have currently been published and how can these applications be classified?} In section \ref{scenarios} we present the three main versions of blockchain applications and we further explain them in \ref{app_cryptocurrency}, \ref{app_smartcontracts}, \ref{app_generalapp}. First, we discuss cryptocurrency including interoperability and briefly, off-chain payment channels. We continue to discuss smart contracts in appendix section \ref{app_smartcontracts} including current security issues and trends. In doing so we provide both practitioners an scholars with trends for these two applications. Finally, appendix section \ref{app_generalapp} shows a wide array of BCT based applications. An interesting trend that can be observed is that almost all of these applications use a form of cryptocurrency to exchange assets and smart contracts to automatize processes. The oversight provides an insight into the rich palette of blockchain applications.

\subsubsection{RQ 3: What are the properties of blockchain technology and what are their trade-offs?} The architecture properties of blockchain technology are discussed in Sec. \ref{charactbct}.
The systematic analysis presented in Sec. \ref{charactbct} couples properties to architectural design. When discussing BCT literature generally posits certain properties. However, the analysis shows that these properties depend on the architectural design of a blockchain. Understanding the implications of certain design decisions is therefore crucial. In section \ref{gtlogicalviewbct} we discuss the relation between BCT software elements and properties.

\subsubsection{RQ 4: What are the challenges for blockchain technology?} The challenges of BCT are presented in section \ref{challoutlook} along with some directions that are currently being pursued to address these challenges. Section \ref{gt:propertiesandchallenges} discusses these challenges and their relative importance to establish a notion of their urgency. The identification of these challenges aid in formulating a future research roadmap and establishes and outlook how they are addressed currently.

In the section hereafter we address the fourth and final research question by presenting future opportunities for research.

\section{Research Gaps and Roadmap}
\label{gapsandfuture}
The field of BCT is rapidly developing yet the results of this study point out four main research gaps that provide opportunities for future research. The research gaps and opportunities for future research have been identified using the results of the GT coding process (Sec. \ref{sec_concepts}) combined with the descriptive (Sec. \ref{sec: results}) and in-depth analysis of the papers under review (Sec. \ref{disc:highlightsandobservations}). 

\subsection{Consensus Protocols}

Despite the fact that several publications discuss consensus protocols, few claims regarding the performance of these protocols are substantiated with evidence. One of the research gaps identified by this study is that evidence is lacking because the consensus protocols suitable for public blockchains remain untested. Similar to the approach presented in \cite{dinh2018untangling} for private blockchains, feature research endeavors could \emph{empirically test} the properties of consensus protocols for public blockchains. Addressing this research gap is especially important since the results of this study suggest that throughput and latency issues are still the major concern, and have a relative high urgency (see Sec. \ref{gt:propertiesandchallenges}). Moreover, these empirical results aid practitioners making more informed architectural design choices. 
There is a trend towards developing consensus protocols based on \emph{sharding} which could be further explored by future research as with the exception of \cite{luu2016sharding} little work on this topic has been published. In the same vein, a shift towards using PoS based consensus protocol for public blockchains can be observed (see Sec. \ref{gt:processviewbct}). However, most published works on consensus protocols only specifically discuss PoW or provide a broad overview (see Sec. \ref{sec: results}). Only three works \cite{kiayias2017ouroboros, pass2017sleepy, li2017securing} have been published on the topic of PoS, and only one of these \cite{li2017securing} addresses the security of PoS protocols. More research on how PoS consensus protocols can effectively ensure the integrity of public blockchains is needed. Especially since there are still open issues to address such as \emph{determining deposits} required for staking \cite{li2017securing} during a PoS based protocol. 

\subsection{Data Storage and Privacy}

Privacy has been shown to be a fundamental issue for BCT (see Sec. \ref{gt:propertiesandchallenges}). As a remedy zk-SNARKs have been proposed (see Sec. \ref{zkp} and Sec. \ref{gt:processviewbct}). It has been claimed zk-SNARKS incur \emph{large overheads} in terms of storage space. However, there is little evidence to substantiate this claim. Future studies could focus on investigating this claim and if proven to be true, investigate manners to decrease this overhead. Although zk-SNARKs facilitate private transactions in our sample no studies were found identified that investigate their effects on performance aspects (throughput and latency). Forthcoming research efforts could determine \emph{the impact of zk-SNARKS on performance} which aids in further developing ZKP protocols in general. There is little work on how blockchains can become made \emph{GDPR compliant} (see Sec. \ref{datapriv}). However, we posit that like any other technology BCT has to operate within the current legal framework. The increasing amount of \emph{data storage space} required by nodes is an unaddressed challenge. Some efforts have been made to address this problem such as the introduction of lightweight clients. The results of our study show that currently data storage is not perceived as a challenge with a high importance (see Sec. \ref{gt:propertiesandchallenges}). Yet we posit that in time the problem might become more urgent when the throughput of consensus protocols will increase and in parallel, the amount of data required to be stored \cite{swan2015blockchain}. Thus, future research should investigate more efficient ways to store data, or determine if data can be omitted from the ledger.

\subsection{Smart Contracts}

From the results of this study \emph{smart contracts} emerge as an important concept related to BCT (see Sec. \ref{sec_concepts}). All of the works in our sample investigates to smart contracts deployed on the Ethereum blockchain platform. Yet, more rigorous, \emph{generalizable}, and \emph{formalized approaches} to analyzing the safe and secure executability of smart contracts is lacking. Novel research could focus its efforts on ensuring the security and safety of smart contracts for platforms other than Ethereum. Literature related electronic contracts (e.g. \cite{marjanovic2001towards, krishna2005contracts, chiu2003three}) could be a potential inspiration how to address the security issues of smart contracts. Furthermore, further work is required to establish the viability and implementation of \emph{legally-enforceable} smart contracts. Given the trend towards sharding the consensus protocol research should be conducted how \emph{cross-shard smart contract validation} can be performed.

\subsection{Usability}
 
This study has shown that multiple applications for BCT have been explored. However, the usability of BCT and smart contracts in particular remains limited since \emph{user-friendly blockchain-oriented tools} are not widely available to non-programmers. The trends in challenges mentioned (see Sec. \ref{gt:propertiesandchallenges}) suggest that over time this will become a more pervasive problem. Providing tools and approaches to ease the development of DApps or smart contract could open up BCT to a broader audience. Therefore \emph{user-friendly tools for smart contract} development would be helpful.

\section{Limitations and Threats to Validity}
\label{Limitations}

Using the guidelines provided by Wohlin et al. \cite{wohlin2012experimentation} the limitations and threats to validity for this study were identified and are discussed in this section. \\

\textbf{External Validity.} First, developments in the field of BCT are introduced at a fast pace. Hence, some of these developments may exist but have not been published yet. Some papers on the topic are so novel that they have not been indexed yet and not included in the selected items. Another issue is that the terminology is still evolving and universal definitions for concepts such as BCT based smart contracts have not yet been formalized. This issue has been addressed by including search terms that are being used interchangeably in the search string (e.g. distributed ledger and blockchain technology) to ensure that all potential aliases of blockchain technology were covered. Items that were found using the search terms have been assessed thoroughly based on various dimensions of quality employing inter-rater reliability. Another threat to validity stems from including aliases of BCT is that findings of the study could also be related to distributed ledger technology. As a strategy to mitigate these threats we set out to define BCT based on a formal concept analysis using literature that specifically mentioned BCT in its introduction and background section. 

\textbf{Internal and Construct Validity.} Second, to attain the results of the research questions a Glaserian-Straussian GT \cite{glaser2017discovery} Grounded-Theory coding approach has been used. Although an inter-rater measurement has been employed, the risk of observer bias is still present. 
Some additional codes were added to the list established during the pilot study. On these codes however, no inter-rater assessment has been performed.
Furthermore, the current body of knowledge on the topic of BCT to day remains limited. Many proposed consensus protocols in the field of BCT have not been rigorously empirically tested in terms of their properties. A prime example being the PoS consensus protocol for which a thorough assessment of the properties (e.g. throughput and latency) of its many variants is lacking in literature. In the same vein, with the notable exception of Li et al. \cite{li2017securing}, not much research has been carried out with regard to the security of the PoS variants.
In addition, in section \ref{processview} we have discussed several consensus protocols found in literature from a process view, indicating their process flow, idiosyncratic security issues and other properties. 
For the sake of space, only the consensus protocols that were recurring more than 3 times in at least 2 different papers across our primary studies were included in this study. However, as mentioned prior, this study has also identified other consensus protocols that were not discussed in full for space sake. Third, the findings of this research are partially based on grey literature sources. Inherent to grey literature is that the quality and accuracy of these sources can be disputable. In order to mitigate this threat we assessed each grey literature item obtained through our search strategy using multiple criteria based on the guidelines provided by Garousi et al. \cite{garousi2017guidelines}. Furthermore, the assessment of the grey literature by the first and second author of this study has been subjected to an inter-rater reliability test (see Sec. \ref{interrater}).

\section{Conclusions}
\label{conclusion}

This study was enacted to (1) analyze how blockchain technology can be defined, (2) provide a systematic overview of the state of the art concepts around that definition, (3) distill a grounded research roadmap around the topic. In section \ref{bctdef}, using a Formal Concept Analysis (FCA) \cite{vskopljanac2014formal} approach, a systematic definition of BCT was distilled.

Beyond the operational definition above, using the well known 4+1 software architecture viewpoint framework \cite{kruchten19954+}, the architecture elements of BCT were fleshed out, specifically, our results recap: (1) the way a platform can be designed: (2) how transactions are processed and, (3) the architectural arrangements typically used for the P2P network underlying BCT. The third aim of this research was to flesh out what blockchain based applications have been discussed in the state of the art and how these can be categorized. The study reveals that there are three types of use cases for BCT: (1) cryptocurrencies, (2) smart contracts and (3) an array of more general applications which can be sub-categorized into five categories, namely, (a) security and fraud detection, (b) securities and insurance, (c) record-keeping, (d) Internet-of-Things (IoT) as well as (e) smart legal contracts. As a fourth objective of this study, we set out to determine the architecture properties of blockchain technology and their trade-offs. Data analysis reveals 8 coupled architectural characteristics --- these properties are \emph{trade-offs} exercised during blockchain architectural design. The fifth objective of this research was to identify the challenges for BCT. In the future, 6 main challenges for blockchain technology need to be addressed: (1) decreasing latency for the conformation of transactions; (2) increasing the throughput of transactions which is related to the design of the consensus protocol; (3) decreasing data storage requirement; (4) protecting the privacy of blockchain users; (5) data governance of blockchain networks; (6) the usability of the technology. 

Finally, an analysis of the papers under review demonstrates that there are four research gaps that need to be addressed by future research concerning: (1) consensus protocols, (2) Data privacy and storage, (3) smart contracts, and (4) the usability of blockchain for end users. In our own research agenda, we plan to further analyse the results and data stemming from our study to further provide architectural and decision-making instruments for practitioners and academics alike. Furthermore, we plan to focus around the social and societal concerns around BCT, namely, its \emph{privacy-by-design} \cite{Guerses2011} aspects as well as its end-user acceptance and maturity.

\begin{acks}
We would like to thank Omkar Khair for his assistance during this research and his valuable practitioners' perspective on BCT. Furthermore, some of the authors' work is supported by the EU H2020 framework programme, grant ``ANITA" under Grant No.: 787061 and grant ``PRoTECT" under Grant No.: 815356.

\end{acks}

\bibliographystyle{ACM-Reference-Format}
\bibliography{sample-bibliography}

\appendix

\begin{screenonly}
\section{In-and exclusion criteria and Replication Package}
\label{replpack}
 All in- and exclusion criteria used for literature, data and resources used in the scope of this study were bundled up and made available for replication purposes. The package in question is available online here: \url{https://drive.google.com/open?id=163pHkUcEHv0tlRKq7MF1pzK0-sM9COGB}
 
\section{Formal Concept Analysis Method}
\label{app_FCA}
\input{appendix_fca.tex}

\section{Grounded-Theory Analysis Method}
\label{app_GT}
\input{appendix_groundedtheory.tex}

\section{Inter-Rater Reliability Assessment Method}
\label{app_interrater}
\input{appendix_interrater.tex}

\section{A 4+1 Overview of Blockchain Technology}
\label{app_4+1}
\input{appendix_4+1.tex}

\section{A Catalogue of Usecases for Blockchain Technology}
\label{app_usecasecat}
\input{appendix_applications.tex}

\end{screenonly}

%% file: appendix_fca.tex
For this research we aim to define BCT from a software architecture perspective based on the extant literature (as stated in research question 1). Hence, the objects used for conducting the FCA were all GL and SL items selected for this study. We first selected the definition of software architecture provided by Bass et al. \cite{bass2003software}:

\begin{quote}
    "The software architecture of a system is the structure or structures of the system, which comprise software elements, the externally visible properties of those elements, and the relationships among them."    
\end{quote}

We have perceived software elements as the main attributes of BCT. The FCA has been conducted in the following steps:

\begin{enumerate}
    \item Each selected item has been reviewed to identify all software elements of BCT (e.g. nodes and block) stated in the introduction or background sections using the codes that were generated during the open coding process of the GT approach.
    \item Whenever a software element was identified it was given a specific code (BCTEL).
    \item All of the identified software elements were identified for the FCA were placed in a matrix. The rows of the matrix represented the GL and SL items whereas the columns represented the software elements that were identified in these items. If one of the items described a particular software element an X was placed in the corresponding cell.
    \item From the analysis a pattern of 10 attributes emerged that were most recurring. A table that lists the most recurring attributes found in the items examined can be found online (see \ref{replpack} for more details). These attributes have been used to construct a definition of BCT (discussed in Section \ref{bctdef}).
\end{enumerate}

%% file: appendix_groundedtheory.tex
Our GT coding process encompassed the following phases:

\begin{enumerate}
\item Open coding - (four phases). During the first phase of the open coding we have conducted a pilot study: Fifteen primary studies form the selected items have been randomly selected to establish an initial set of codes using open coding. A second pass has been made on the pilot papers to generate a final list of codes, in order to minimize inconsistencies during the coding process. In the second phase, using the priorly established set of codes, an initial theory about the relation between codes has been developed based on the pilot study. The third phase consists of constant comparison: the pilot study initially generated 266 codes. Next, these initial codes were organized into a hierarchy of codes based on emerging relations between concepts. The resulting structured start-up list of codes was used to code the remainder of the primary papers. The coding has been executed in parallel by two different coders, over two equally divided splits, to ensure avoidance of observer bias. Each primary paper has been analyzed line by line with the list of codes. Codes were applied if they reflected a concept in a paragraph. For example, the paragraph: \emph{"An important component of the blockchain technology is the use of cryptographic hash functions for many operations, such as hashing the content of a block. Hashing is a method of calculating a relatively unique fixed-size output (called a message digest, or just digest) for an input of nearly any size (e.g., a file, some text, or an image)."} would be coded with the code ``FUNC-HASH", where FUNC denotes the code's reference to functional aspects of blockchains while the -HASH refers to a refinement of such functional aspects to refer to hash-functions, in the specific. The fourth phase, constant memoing, has been conducted simultaneously with the third phase. During this phase we have kept notes to capture key messages, relations and observations on the analysed texts. \\
\item Axial coding - (two phases). In the first phase comparing the concepts coded has led us to inductively generate relations among coded concepts (e.g. between consensus protocols and blockchain challenges); For the second phase, the definitions of all concepts coded have been compared with each other to identify aliases.\\
\item Selective coding - (three phases). The third step involved the selection and arrangement of codes to form a relationship model. As a first phase have we arranged the data: Every portion of text that was coded with a certain code has been placed in a table. Each of these codes represented a core concept observed in the literature (e.g. consensus protocol, blockchain networks and applications). Next, in the second phase we have modeled the data. The data was represented in a view consisting of several diagrams. Whenever possible these diagrams were connected to one another, resulting in the construction of the views of BCT architecture used later in Sec. \ref{sec: results} to flesh out results. Finally, as part of this phase, the diagrams and all the data at hand has been analyzed and sorted to address the research questions behind this study.\\
\end{enumerate}

%% file: appendix_interrater.tex
The first two authors of this paper examined their inter-rater reliability via the following approach: First, the inter-rater reliability of the assessment of the SL items was determined. The Krippendorff test revealed that there was 86\% inter-rater agreement ($\alpha$ 86\%). Second, another Krippendorff $\alpha$ test was employed to determine the inter-rater reliability between both observers with regard to the in- and exclusion criteria for GL. The resulting K-$\alpha$ statistic ($\alpha$ 93\%) showed that there was a high inter-rater agreement on the exclusion and inclusion of GL items. Third, an inter-rater reliability of the analysis that was conducted concerning the quality of the GL studies was tested using a Krippendorff $\alpha$ test. The results of the test indicated that there was a 99\% agreement ($\alpha$ 99\%) between both observers. Finally, the findings presented in the sections hereafter were attained by adopting a thematic coding process. Therefore as a precautionary measure to avoid observer bias during the coding process we conducted an inter rater reliability assessment test to the codes that were obtained from the pilot study. The results of the Krippendorff's $\alpha$ test suggest that there was a 69\% agreement between the observers. Because this result was bellow the commonly accepted threshold of an $\alpha$ of 80\%, the first two authors deliberated over the differences to form one consistent initial coding set.

%% file: appendix_4+1.tex
\subsection{Logical View}
\label{app_logicalview}

The logical view emphasizes on the functional requirements and services the system should provide to its end users \cite{kruchten19954+}. Decomposition of the architecture aids in identifying the elements that are common across the system. We used an ontology for BCT as proposed in \cite{de2017understanding, glaser2017pervasive} to organize the discussion of its main elements.

\subsubsection{Transactions}
Transactions in a blockchain system are executed using \emph{public key cryptography}. Actors that send transactions between themselves and other beneficiary actors. Two pairs of keys are used to allow actors to interact with one another; a private and a public key that are mathematically related to each other \cite{azouvi2017secure, internetofthingschristidis2016blockchains, yaga2018blockchainoverview,aste2017blockchain}.

Actors can use their secret \emph{private key} to sign transactions, that are addressable on the network via their public key. Private keys are stored in software called a \emph{wallet} that is installed on a hardware device. A wallet can also store public keys and associated addresses to send transactions to \cite{yaga2018blockchainoverview, DistributedLedgerTechnologyandBlockchain2017, yin2018anti}. The \emph{public key} is widely disseminated without reducing the security of the transaction process \cite{yaga2018blockchainoverview, DistributedLedgerTechnologyCybersecurity2016}. Public keys have various functions they are used to derive addresses and to verify the signatures generated with the private keys \cite{distributedledgertechnologyinpayments2016, yaga2018blockchainoverview} (e.g., the Elliptic Curve Digital Signature Algorithm (ECDSA) \cite{mingxiao2017review,yin2018anti,van2017process}). A transaction transfers an amount of coins from one \emph{input address} owned by the private key owner to one or several \emph{output addresses} \cite{ruffing2014coinshuffle}. An \emph{address} is created using the public key, the private key, and a cryptographic hash function \cite{decker2015making}.   

\emph{Cryptographic Hash functions} are used for the purpose of many operations, such as signing transactions (SHA-256 in the Bitcoin case\cite{nakamoto2008bitcoin}). \emph{Hashing} is the process of converting any data of arbitrary size to data of fixed size where the output is a bit-string known as the digest, hash value, hash code or hash sum \cite{yaga2018blockchainoverview, peters20166understanding,deblockchainsforgovernmental2017distributed, DistributedLedgerTechnologyandBlockchain2017}. Tampering with the original transaction data would immediately get noticed as the hash would differ from that previously generated and recorded on the blockchain \cite{deblockchainsforgovernmental2017distributed, olnes2017blockchain, DistributedLedgerTechnologyandBlockchain2017,aste2017blockchain}. Hash algorithms are generally designed to be \emph{one-way}, meaning that one cannot compute and extract original data from a hash \cite{yaga2018blockchainoverview, deblockchainsforgovernmental2017distributed, distributedledgertechnologyforthefinancialindustry2016,aste2017blockchain}.

\subsubsection{Nodes}
The peers in the P2P network, also referred to as nodes are devices capable of processing and verifying transactions. Depending on the permissions all nodes or a specific subset of nodes validate transactions. The \emph{permissions}, refer to the rights that are granted to the nodes in a blockchain network. There are three major types of permissions \cite{globalblockchain2017}: \emph{Read permission} that dictates which nodes can access the ledger and audit transactions, \emph{Write permission} that stipulates who can create transactions and broadcast them to the network, and \emph{Commit permission} that describe who can update the ledger. Some nodes may only have permission to use the services in a blockchain network such as announcing an transaction, while other nodes are permitted to propose to include certain unspent transactions in a block to update the ledger \cite{DoyouneedaBlockchain?2017, centralbankofbrazil2017distributed, pirlea2018mechanising}. Other blockchains (e.g. MultiChain \cite{MultiChainPrivateBlockchain2015}) allow for more fine grained permissions, for example the \textit{permission to create assets} \cite{arevolutionintrust2017, yaga2018blockchainoverview, globalblockchain2017, xu2017taxonomy}.

Updates and changes to the software of a blockchain are called \emph{forks} \cite{yaga2018blockchainoverview, DistributedLedgerTechnologyandBlockchain2017}. Recall from section \ref{Basic Blockchain Technology Terminology} that during the mining process two miners might propose a block at the simultaneously. This type of fork can be described as an \emph{accidental fork} that stems from the probabilistic nature of the mining process \cite{DistributedLedgerTechnologyandBlockchain2017}. To resolve the fork the miners as a rule follow the longest chain. The longest chain rule can also effectively be employed to invoke two other types of forks and propose updates to the blockchain \cite{DistributedLedgerTechnologyandBlockchain2017}. The first of these types of forks is a \emph{hard fork}. 

A hard fork brings about updates to the BCT that prevents nodes that do not accept the fork from using the changed BCT or the nodes can continue to use the original protocol without the update. Nodes that use dissimilar hard forks cannot interact. Structural changes to a blockchain requires hard fork \cite{yaga2018blockchainoverview} and can also effectively create a new blockchain \cite{xu2017taxonomy}. 

The second type of fork is a \emph{soft fork}. Only a super majority of nodes need to upgrade to implement the novel rules stipulated by the soft fork. Nodes that did not update still accept newly created blocks as valid after the soft fork \cite{back2014enabling, yaga2018blockchainoverview}. At a later point in time the two chains are reconciled with one another. The possibility to make updates to a blockchain or distributed ledger implies that there are also \textit{Update permissions}.

There exists at least three categories of blockchain networks \cite{xu2017taxonomy,Zheng2017AnOverviewofBlockchainTechnologyArchitectureConsensusandFuturetrends}; Public, private, and consortium networks that have different arrangements in terms of their permissions. A node that has permissions and function to verify transactions that are executed on a private or consortium blockchains is a \emph{validator} \cite{realizingthepotentialof2017,distributedledgertechnologyforthefinancialindustry2016,BlockChainTechnologyBeyondBitcoin2015,cachin2017blockchains, de2017understanding}. On a public network a node can also validate transactions by proposing blocks \cite{dinh2018untangling}. Nodes that fulfill the aforementioned are also called \textit{miners} or \emph{forgers}. Miners or forgers have two separate functions; (a) to correctly construct and propose new blocks, and (b) to check the validity of the transactions in each block \cite{luu2015demystifying}. In the specific case of some blockchain platforms (e.g. Ethereum) that allows for the creation of smart contracts, miners, forgers and validators have the additional task of executing the smart contracts to check for the validity of their outcomes.

\subsubsection{Blocks}

On a blockchain transactions are stored in \emph{blocks}. Blocks can be divided into two parts \cite{yu2018virtualization, mingxiao2017review, yin2018anti}; (1) a \emph{blockbody} that contains the verified transaction data, which is recorded in the form of a Merkle tree \cite{yu2018virtualization, xu2017taxonomy}, a Patricia Merkle tree \cite{dinh2018untangling, glaser2017discovery} or Bucket hash tree \cite{dinh2018untangling}, and (2) a \emph{blockheader} that specifies the elements required to guarantee safety. Rather than storing the hash of each transaction individually these are stored in a data structure known as a Merkle tree to further diminish the storage requirements. 

A Merkle tree combines the hash values of the transaction data by re-hashing them until there is a single root left which is called the \emph{root hash} \cite{nakamoto2008bitcoin,yaga2018blockchainoverview, Zheng2017AnOverviewofBlockchainTechnologyArchitectureConsensusandFuturetrends, dinh2018untangling}. Since the root hash is a hash function it can be used as a mechanism to summarize transactions that have been stored in a block, any alterations to the underlying transactions would be detected \cite{yaga2018blockchainoverview, glaser2017pervasive,xu2017taxonomy}. Some blockchains employ other means of data structuring that also facilitate in capturing smart contract states besides transactions. For instance, Ethereum employs a Patricia merkle tree whose leaves record key-value states \cite{dinh2018untangling}. Hyperledger implements a Bucket-Merkle tree that groups states into a pre-defined number of buckets \cite{dinh2017blockbench}. The maximum number of transactions that a block can contain depends on the block size and the transaction size  \cite{mingxiao2017review}. The blockheader encompasses six elements \cite{DistributedLedgerTechnologyCybersecurity2016, Zheng2017AnOverviewofBlockchainTechnologyArchitectureConsensusandFuturetrends, yu2018virtualization,mingxiao2017review}; (1) a root hash, (2) a block version number that specifies the software version of the block, (3) Blockheader hash of the previous block, (4) the timestamp of the block, (5) the difficulty (target) required to create the block and;(6) a nonce random number.

\subsubsection{Chains}
Each block is linked to its predecessor known as \emph{parent block} by including its blockheader hash to form an integral chain of blocks that can be traced back to the first, or \emph{genesis block}. Hence the term "blockchain" technology. By hashing transactions and chaining all blocks to one another a blockchain provides a data model that allows to track all historical changes to the distributed ledger. Moreover, by combining the hashing of transactions and the chaining of blocks the distributed ledger becomes tamper proof\footnote{According to Stark \cite{applicationsofdistributed2017} this term needs to be qualified because there is a very high probabilistic guarantee that the data recorded on the blockchain is not changed.}. 

Novel blocks are generated using a consensus protocol. Provided that the transactions included in the newly proposed block are valid, each new block enhances the security guarantees of the block before it \cite{DistributedLedgerTechnologyandBlockchain2017,nakamoto2008bitcoin,mingxiao2017review}. The number of blocks in the chain between the last created block and the genesis block is called the \emph{block height} (So a genesis block has height 0.). Because of the probabilistic nature of some consensus protocols (e.g. PoW and PoS) two miners can propose a valid block simultaneously; This can result in the situation where \emph{stale blocks} are created which will never be included in the longest chain, and can therefore be considered as wasted efforts \cite{luu2015demystifying}. The Ethereum platform refers to these blocks as \emph{uncle blocks} \cite{EthereumHomesteadDocumentation2017,Zheng2017AnOverviewofBlockchainTechnologyArchitectureConsensusandFuturetrends}. Forks that occur accidentally or are invoked with ill intend are resolved using a \emph{fork choice rule function} \cite{pirlea2018mechanising} (e.g. longest chain rule). Eventually, one of these chains will become the longest chain and the other shorter chains will be abandoned. Blocks that are included in the abandoned are colloquially known as \emph{orphan blocks} \cite{centralbankofbrazil2017distributed, pirlea2018mechanising}.

\subsection{Development View}
\label{app_developmentview}
The deployment viewpoint defines how the various elements identified in the logical, process, and implementation viewpoints mapped onto the various nodes \cite{kruchten19954+}. 

Every node in a blockchain network has two layers; an application layer and a blockchain layer \cite{xu2016connectblockchain}. Therefore, there are two groups of developers to be considered: First, Developers utilizing the services of a blockchain platforms (e.g. Bitcoin or Ethereum) to develop decentralized applications. Second, developers that seek to create a new blockchain platform.

\subsubsection{DApp developers}

The blockchain networks can be leveraged to build \emph{decentralized applications} (DApps) that use their services. DApps are applications designed have a distributed nature as they are run on a P2P network instead of one computer. At first sight a DApp looks similar to that of a normal (web) application. Instead of using an API to connect to a database, a smart contract will connect the DApp to the blockchain for all required information. The front end of a DApp can therefore be regarded as a facade that allow users to interact with the services provided by the blockchain platform. Using the smart contract services of blockchain networks such as Ethereum allows developers to create custom DApps suitable for a wider range of applications. Developing a DApp might include coding a smart contract that needs to be compiled using a virtual machine that is offered by a blockchain platform. The final step is to deploy the smart contract on the blockchain network by using a wallet. Once the final step has taken place the DApp can interact with the blockchain platform by invoking smart contracts. 

\subsubsection{Establishing the P2P network}

Developers seeking to build their own blockchain platform have to program multiple software packages. Enabling transactions forms the basis for any blockchain network. A wallet needs to be programmed to allow clients of the platform to interact with other peers in the network. In order for peers on the networks to interact they need to be knowledgeable about one another. Therefore there should be a \emph{address propagation} method installed. Next the nodes need to connect via \emph{peer discovery}. Peers of the Bitcoin network for instance, connect to each other over an unencrypted TCP channel. Every node keeps a list of IP addresses associated with its connections since there is no authentication of nodes. However, for networks that consist of peers that already know each other a different approach to peer discovery might be possible. Another aspect is the mechanism for \emph{propagating transactions} \cite{biryukov2014deanonymisation}. The Bitcoin uses a propagation mechanism where the nodes in the P2P network forward transactions to their neighbors (known nodes). Other platforms such as Ripple \cite{schwartz2014ripple} use a pre-defined node list that a node must store to process transactions.

\subsubsection{Data Model}

Data with regard to transactions can be stored in two ways: As a first method, like the Bitcoin, one can choose to add data into transactions. Another second method is to add data into contract storage like Ethereum \cite{xu2016connectblockchain}. The state of the distributed ledger is a collection of all the accounts that have not been spend yet, referred to as the \emph{unspent transaction output model} (UTxO) \cite{DistributedLedgerTechnologyCybersecurity2016,Zheng2017AnOverviewofBlockchainTechnologyArchitectureConsensusandFuturetrends}. The Bitcoin can therefore be regarded as having a transaction based model. Another choice that could be made is to offer smart contracts as a service.
In these instances the transactions resulting from the execution of smart contracts need to be stored. Ethereum stores smart contracts in specific accounts using an \emph{account based data model} \cite{dinh2018untangling}. The state of transactions in the system are the changes of the complete contract storage expressed as key-values \cite{xu2016connectblockchain}. Other platforms such as Hyperledger simply use key-values to store data in Docker containers \cite{dinh2018untangling}. 

\subsubsection{Consensus protocol}

A consensus protocol can be regarded as a sequential set of steps that stipulate the rules of engagement for the network to process transactions and smart contracts. Various consensus protocols are available for this purpose with idiosyncratic properties in terms of throughput, latency and security. However, developers can also opt to design a novel consensus protocol from scratch. To verify transactions during a consensus protocol the wallet should be able to import a copy of the distributed ledger. To compile the smart contracts written by users of the platform into bytecode that can executed by the platform, the creation of a virtual machine might be needed. This virtual machine is also used to verify the outcomes of the execution of smart contracts. Platforms like Kadena use an interpreter language to write smart contracts which makes a virtual machine redundant. Permissions should be allocated and distributed among nodes to confine permissions of nodes to participate in the consensus protocol if desired. A mechanism to distribute economic incentives need to be devised for nodes that verify transactions \cite{yu2018virtualization,xu2016connectblockchain, mingxiao2017review}.

\subsection{Process view}
\label{app_processview}

The process view specifies which thread of control execute the operations of the classes identified in the logical view. It also takes into account non-functional requirements such as performance and system availability. The consensus protocol is at the heart of all BCT processes since it allows for the enactment of transactions and ensures that the distributed ledger remains consistent. 
Consensus protocols are discussed in chronological order.

\subsubsection{Practical Byzantine Fault Tolerance}(PBFT).
PBFT is mostly used in a private setting for permissioned blockchains because it assumes authenticated nodes \cite{dinh2018untangling, Zheng2017AnOverviewofBlockchainTechnologyArchitectureConsensusandFuturetrends, xu2017taxonomy}. The protocol itself is exclusively based on communication, and nodes go engage in multiple rounds of communication to reach consensus \cite{dinh2018untangling}. Nodes do not get a reward for achieving consensus, rather in the event of malicious behavior by an authenticated node it can be held legally accountable \cite{peters20166understanding, globalblockchain2017}. A primary leader node mines the blocks. The leader can be changed by other nodes via a "view-change" voting protocol, in the occurrence of a crash or when it exhibits malicious behavior \cite{internetofthingschristidis2016blockchains, mingxiao2017review}. PBFT as has five phases to reach consensus \cite{mingxiao2017review}:

\begin{enumerate}
\item \textit{Request}. A client sends a request for a transaction to a leader node, that accordingly gives the request a timestamp.
\item \textit{Pre-prepare}. The leader node records the request message and renders an order number for it. Then the leader node broadcasts a pre-prepare message with a value to the other nodes. Initially the nodes decide on whether to accept the request or to reject it.
\item \textit{Prepare}. When a node decides to accept the request it broadcasts its values in the form of a message to other nodes. Nodes then receive prepare messages from one another. Once a node has collected sufficiently enough messages from other nodes (2f+1 messages), hence if a majority of nodes decides to accept the request, it will enter the commit phase.
\item \textit{Commit}. All nodes that are involved in the commit state send a commit message to the other nodes. Concurrently, if a node receives a message of acceptance from 2/3 of the other nodes in the network a consensus has been reached to accept the request. Next, the node executes the instructions that are stated in the request message.
\item \textit{Reply}. The nodes in the network reply to a request of a client. If a delay in the network occurs and the client did not receive a reply message, the request will be resend. In the case that the request did get executed the nodes in the network need to send a reply message repeatedly.
\end{enumerate}

The protocol require nodes to have a high degree of trust and is not suitable for permissionless blockchain. PBFT is purely communication based rather than on cryptographic primitives. Therefore, whenever an authenticated peer is attacked, the blockchain could easily be manipulated \cite{deblockchainsforgovernmental2017distributed}. PBFT does not scale well as executing the PBFT consensus protocol can be time consuming. As more nodes join the network the latency increases and throughput decreases \cite{mingxiao2017review}.

\subsubsection{Proof-of-Work}\label{app_pow}(PoW) is often referred to as the \emph{Nakamoto consensus protocol} \cite{luu2016making, voyiatzis2017merged, luu2015demystifying,pass2017sleepy}. Recent estimations indicate that the majority of public blockchains use PoW as their mechanism to create a consensus \cite{dinh2017blockbench,deblockchainsforgovernmental2017distributed}. Public blockchains need to have a high degree of Byzantine fault tolerance as users can not trust one another. The PoW consensus protocol is designed for the case where there is little to no trust amongst users of the system \cite{yaga2018blockchainoverview}. Furthermore, it safeguards against Sybil attacks which are common in open, decentralized environments in which a malicious actor can acquire multiple identities \cite{dinh2017blockbench}. 

Consensus in PoW is achieved through a hashing competition between miners. Competing miners need to commit computing power to calculate the solution to the same mathematical problem. To incentivize miners to participate in the consensus process the miner that is the first to find the solution to the mathematical problem reserves the right to publish the next block, and is rewarded by an amount of cryptocurrency\footnote{In the case of Bitcoin the amount of cryptocurrency that is provided to miners for their services decreases to zero over time as it halved every 210,000 blocks \cite{internetofthingschristidis2016blockchains}. From that point on, miners can only be compensated by transaction fees alone \cite{nakamoto2008bitcoin}} \cite{mingxiao2017review, yaga2018blockchainoverview, howblockchain, xu2017taxonomy}. In addition, the miner to win the competition with its peers is also be able to collect the transactions fees that were paid by clients\footnote{Note that transaction fees are not obligated to execute Bitcoin transactions but rather have been paid by clients to prioritize their transactions \cite{realizingthepotentialof2017}.}. 

Finding the solution to a PoW problem is a computationally arduous process for which there are no shortcuts \cite{DistributedLedgerTechnologyandBlockchain2017, yaga2018blockchainoverview}. The solution to the problem is hard to find, yet easy to check once they have been found \cite{luu2015demystifying}. Given that only one miner can win the competition and is rewarded the other nodes have simply wasted resources (CPU power and energy) in their attempt \cite{swan2015blockchain,yaga2018blockchainoverview,gatteschi2018blockchain,mingxiao2017review,xu2017taxonomy}. In addition, because the difficulty of PoW problems increases over time  makes it even harder to win the competition \cite{yaga2018blockchainoverview}. Some miners have as a result started collaborations called \textit{mining pools} to address this problem. When participating in mining pools miners combine their computing power and divide the work to find the solution \cite{realizingthepotentialof2017}. The rewards for proposing the next block are split based on each node's contribution \cite{voyiatzis2017merged}. However, pooled mining could open the opportunity for a 51\% attack if a cartel of large mining pools were to control more than 50\% of the hashrate  \cite{mingxiao2017review,realizingthepotentialof2017}. The large accumulation and \emph{centralization of hashing power} could constitute to a majority vote of miners which could endanger the integrity of the ledger. The steps in the mining process are as follows \cite{garay2017bitcoinbackbone,garay2015bitcoinanalysis}:

\begin{enumerate}

\item \textit{Get the difficulty}. The difficulty of the PoW problem changes every \emph{epoch} (fixed number of blocks) and depends on the \emph{target} that is generated for each block \cite{mingxiao2017review,yaga2018blockchainoverview,garay2017bitcoinbackbone,zohar2015bitcoin}. Therefore, a miner first needs to use a \textit{target recalculation algorithm} to determine the correct target for the problem, which depends on the hash rate of the whole network. The \emph{hash rate} is the number attempts required to solve the PoW \cite{meshkov2017short, mingxiao2017review}. 

\item \textit{Collecting Unspent Transactions}. All requested transactions that have not been processed since the generation of the last block are stalled in a \emph{transaction pool} \cite{yaga2018blockchainoverview, pirlea2018mechanising}. During mining, miners collect and verify all of the transactions that are kept in the transaction pool and hash them to a root hash. To mine the block the miner also needs to include the block version number, blockheader hash of the previous block, nonce random number and the target found during step 1 \cite{mingxiao2017review}.

\item \textit{Calculating the blockheader hash}. Next miners start their attempts to solve the mathematical problem which comprises finding the correct nonce so that the \emph{double hash value of the blockheader} (SHA-256\(^2\)) is equal to or smaller than the number of zeroes (target) that the network demands \cite{internetofthingschristidis2016blockchains, mingxiao2017review}.  

\item \textit{Broadcasting the Block}. If a miner has found the solution it is propagated along with a \emph{candidate block} to the remaining nodes in the network.

\item \label{step proof of work,verification} \textit{Verification of the Block}. Upon receiving a candidate block other nodes will check (a) whether the candidate block includes correct hash references to the previous block on their chain and thus, the proposed nonce to solve the problem \cite{DistributedLedgerTechnologyandBlockchain2017}. Further, (b) whether the block contains valid transactions by verifying the root hash \cite{yaga2018blockchainoverview}. If the block is found to be valid nodes express their acceptance by appending the block to their copy of the blockchain, and resend the block to other nodes. Finally, when a consensus is reached by a majority of nodes in the network (51\%) that the block is valid it will be included in the chain of blocks.

\end{enumerate}

Whenever a fork occurs miners will wait for longest chain to be formed by the addition of new blocks at the cost of high latency. The system is therefore designed under the presumption of \emph{eventual} reconciliation of the ledgers, which can be regarded as non deterministic \cite{dinh2018untangling}. PoW ensures a high degree security as theoretically it has a byzantine fault tolerance of 51\% \cite{dinh2018untangling,mingxiao2017review,gervais2016securitypow}. Some scenarios to attack for PoW attacks have been identified such as a \emph{selfish mining strategy}. In this scenario when miners mine a correct block they do not propagate it. Rather, they privately mine blocks and hide them. The private branch of the chain will only be revealed to other nodes under the right circumstances during which it can result in a higher of frequency of forks that pose a threat to the blockchains security \cite{garay2015bitcoinanalysis, mingxiao2017review,gervais2016securitypow}. Another attack scenario is the verifiers dilemma \cite{luu2015demystifying} wherefore miners do not receive any reward for validating the correctness of a block proposed by another miner. This is because it is assumed that the other miners want to maintain a correct blockchain so that the value of the Bitcoin does not depreciate. However, this assumption creates a dilemma for miners whether or not to actually check the validity of the transactions included in the block as they do not receive any reward for doing so. 

Most PoW consensus protocol variants have a low throughput \cite{sompolinsky2015ghost}. This problem is further amplified because PoW based blockchains also has a relatively high latency (time it takes to confirm the transactions) due to the fact that the block time interval is set at a fixed. This is because between the propagation of two blocks no transactions are being processed. 

\subsubsection{Proof-of-Elapsed Time}PoET is designed to address the inefficiency of PoW and replaces it with a protocol that is based on \emph{trusted hardware}. A node that uses trusted hardware however, can be checked for certain properties such as whether it is running a certain software. This aids in relaxing the trust model in settings were the Byzantine's Generals Problem might be present \cite{dinh2018untangling}. 
 
Sawtooth Lake, a project by Hyperledger, leverages Intel's \emph{Software Guard Extensions} (SGX) to establish a validation lottery that makes use of their CPUs capability to render a timestamp that is cryptographically singed by the hardware \cite{DistributedLedgerTechnologyCybersecurity2016}. The PoET consensus protocol used for Hyperledger Sawtooth Lake is carried out in the following steps;

\begin{enumerate}
\item \textit{Requesting wait time}. Every potential validator node request a secure \emph{waiting time} from their hardware \emph{enclave} that distributes these waiting times randomly. Then every node waits accordingly \cite{yaga2018blockchainoverview, meng2018intrusion}. The trusted hardware can afterwards produce certificates that indicate how much time has expired since the timer has started \cite{dinh2018untangling}.
\item \textit{Election of Validation Leader}
After the assigned time has passed all nodes declare themselves to be validation leaders. Whichever node has the shortest wait time for a particular transaction block is elected the leader \cite{DistributedLedgerTechnologyCybersecurity2016, dinh2018untangling}. More specifically, the node includes its PoET timestamp in the block, and if its waiting time is smaller than that generated by other nodes then the block is accepted \cite{dinh2018untangling}. Because each node has an equal chance of being chosen via their trusted hardware, the probability for a single entity of controlling the validation leadership is proportional to the resources contributed to the overall network \cite{meng2018intrusion, cachin2017blockchains}. 
\end{enumerate}

The steps that follow are similar to that of PoW; the elected leader creates a block and broadcasts it, thereafter the other miners validate the block. The PoET protocol can be regarded as more environmental friendly. However, the probability of becoming validation leader is proportional to the number of trusted hardware modules, and therefore economic investments still enhances one's influence on the consensus protocol. Furthermore, the security of the protocol is dependent on the hardware that could be running on a potential malicious host. The SGX for instance, is susceptible to \emph{rollback attacks} \cite{brandenburger2017rollback} wherein a malicious user provides stale data to the trusted hardware's enclave and \emph{key extraction attacks} \cite{chen2017poet} where an attacker leverages a vulnerability to extract attestation keys from SGX processors enabling a full break of the SGX \cite{cachin2017blockchains}.

\subsubsection{Proof-of-Stake}\label{app_PoS}As a response to the limitations of PoW the BCT community has turned towards Proof-of-Stake (PoS). The PoS consensus protocol has been introduced for public settings \cite{mingxiao2017review} with the aim to safeguard against Sybil attacks and malicious behavior by untrusted nodes \cite{dinh2018untangling}. The PoS protocol offers a more efficient and environmental friendly alternative to PoW as computing power is partially substituted by virtual resources (e.g. cryptocurrencies) that miners must invest to propose blocks \cite{li2017proof, swan2015blockchain, gatteschi2018blockchain, yu2018virtualization}. Rather than using computer power as a scarce resource to generate security, Proof of Stake uses the scarcity of the coin itself. Therefore nodes that participate in a PoS consensus protocol are more commonly referred to as \emph{forgers}  instead of miners \cite{baldimtsi2016indistinguishable, li2017securing}. 

The idea behind the PoS model is that the more assets (e.g. cryptocurrency), or \emph{stake} a node has, its incentive to undermine the system diminishes because subverting the system would inherently mean that the worth of the nodes' stake would decrease \cite{yaga2018blockchainoverview, meng2018intrusion}. Logically, this implies that one cannot participate in the consensus protocol without owning a stake \cite{gatteschi2018blockchain}. A shared commonality of all PoS variants is that nodes that have more stake have a higher chance of generating new blocks \cite{yaga2018blockchainoverview, li2017proof, mingxiao2017review, Zheng2017AnOverviewofBlockchainTechnologyArchitectureConsensusandFuturetrends}. In other words, the more skin a forger puts in the game the higher its reward will be.

\emph{Coin age based PoS} is one of the earliest implementations of the PoS consensus protocol used by PeerCoin. \emph{Coin age} means that besides its market value a currency has an age \cite{Zheng2017AnOverviewofBlockchainTechnologyArchitectureConsensusandFuturetrends, mingxiao2017review}. The accumulated \emph{time} that a node holds his stake (currency) before using them to generate a block \cite{li2017securing}. When used to generate a block the age of a coin is reset to 0. Only after a pre-determined time the coin can be used again as a stake to create a new block. The coin age system introduces the problem of \emph{coin hoarding} which means that nodes do not spend their coins because they incentivized to hold on to them \cite{mingxiao2017review}. Rather than punishing inactive nodes, the Proof of Activity (PoA) protocol rewards stakeholders that contribute in sustaining the network. For stakeholders to collect a block reward they need to spend active online time on the network \cite{bentov2014proofact}.

Cryptocurrencies like Blackcoin and NXT use randomization to predict the next forger of a block and do not rely on a mining process to create blocks. For these \emph{chain-based PoS variants} an algorithm will review all users that have a stake and make a selection of the validating nodes based on the nodes' stake ratio compared to the entire system \cite{yaga2018blockchainoverview}. Although both variant of PoS to a degree succeed in decreasing energy waste, lowering latency and increasing throughput, Li et al. \cite{li2017securing} suggest that both coin age based and the chain based variant of PoS have three security problems: 

The first problem that corrodes the security is the \emph{nothing at stake problem}. This attack can be carried out because they only need to have a stake, but not to commit a partition of the stake for the sake of forging a block. 
A second problem is a \emph{stake grinding attack} were an attacker tries to bias the randomness of the election to forge a new block in their favor. This attack can be mounted in two ways; Either the attacker can "grind" through many combinations of parameters and find favorable parameters that would increase the probability of their coins generating a valid block. Or, the attacker can skip an opportunity to create a block to provide the forger a greater opportunity to forge future blocks in next rounds. The third problem is the \emph{long range problem}, sometimes referred to as a history attack.  
Some blockchain platforms secure their chains for long range attacks by implementing \emph{checkpoints}. A checkpoint is a block after which all prior chained blocks are regarded as final and immutable and are hard-coded in the system's software \cite{li2017securing, deblockchainsforgovernmental2017distributed}. Snow White based upon the work of Pass and Shi \cite{pass2017sleepy} designed to maintain robustness even if nodes sporadically participate in the consensus protocol. Ouroboros protocol \cite{kiayias2017ouroboros}, proposes a novel rewarding mechanism with the aim of increasing the incentive for nodes to behave honestly and to diminish the possibility of grinding attacks.

\emph{Byzantine fault tolerance PoS} is a hybrid consensus protocol that combines the use of stakes (PoS) and aspects of the PBFT consensus protocol \cite{xu2017taxonomy}. For instance, Tendermint adds a multi round voting system which makes the consensus protocol more complex. This PoS variant allows all staked nodes to participate in the block selection process \cite{yaga2018blockchainoverview}:

\begin{enumerate}
\item \textit{Propose}. An algorithm embedded in the blockchain platform will randomly select several nodes with a stake to propose a block in a round-robin fashion. 
\item \textit{Prevote}. All nodes with a stake are asked to cast their vote for the next block. Only when 2/3 of all nodes vote to confirm the block can it continue to the precommit step
\item \textit{Precomit}. In the Precommit stage nodes engage in another round of voting for which yet again 2/3 of the nodes need to confirm the block to proceed to the next step.
\item \textit{Commit} Once all votes have been casted and the block has been accepted it will be appended to the blockchain. A reward will be provided to the proposer of the block.
\end{enumerate}

To overcome the nothing at stake problem the forgers are required to submit a \emph{deposit} in the form of a stake to validate blocks. If forgers on purpose, or otherwise vote for blocks that contain invalid transactions their stake is slashed away and destroyed \cite{DistributedLedgerTechnologyCybersecurity2016, dinh2018untangling}. Deposit based blockchain platforms have the problem that they lock down currency and thus decrease the liquidity of the network's currency. Furthermore, there could be an incentive to propagate conflicting blocks in order to double spend if the value of these transactions surpass the value of the deposit made by the forger \cite{li2017securing}. 

According to literature there is also some general fundamental critique on the design principles of PoS consensus protocols. Firstly, nodes with a higher stake have more chance of mining a block \cite{yaga2018blockchainoverview, mingxiao2017review}. Furthermore, in PoS less energy is wasted and throughput and latency benefits are gained as compared to PoW. Yet, in essence PoS did not completely solve the problem of resource wasting, concentration of hashing power besides low throughput, and high latency. However, as the mining costs are low attacks might come as a consequence which can reduce the security \cite{mingxiao2017review}.

\subsubsection{Delegated-Proof-of-Stake}(DPoS) Delegated Proof-of-Stake introduces another variant of PoS \cite{mingxiao2017review, Zheng2017AnOverviewofBlockchainTechnologyArchitectureConsensusandFuturetrends}. In DPoS stakeholders elect delegates, referred to as \emph{witnesses} to forge and validate blocks in round-robin fashion \cite{li2017securing}. The consensus protocol of BitShares is an example of DPoS that is divided into two parts: electing a group of block producers and scheduling production;

\begin{enumerate}
\item \textit{Vote for Witness}. Each stakeholder can select a witness based on their stake. The top number of the most selected witnesses that have participated in the voting round gain the authority to forge a block.

\item Thereafter the elected witnesses sequentially forge new blocks. A prerequisite is that the witness spends sufficiently enough time online. In the case a witness is unable to forge its assigned block stakeholders will vote for a new witness to substitute the defaulting witness. Accordingly, the activities for that block will be moved to the next \cite{mingxiao2017review}. 

\end{enumerate}

Compared to PoW and Pos, DPoS is more energy efficient. Further, because the voting about the validity of a block is delegated and fewer nodes are needed to validate the blocks can be confirmed more quickly. Hence, as compared to PoW and PoS, DPoS has a low latency. Moreover, parameters including block size and block intervals can be adjusted by \emph{committee members} of the governance board. When a delegate acts malicious this dishonest delegate can be voted out by all the other nodes \cite{Zheng2017AnOverviewofBlockchainTechnologyArchitectureConsensusandFuturetrends,li2017securing}.

\subsubsection{Zero-Knowledge-Proofs}
\label{app_zkp}

Recently, different Zero-Knowledge-Proofs (ZKP's) based BCT networks have been proposed to preserve users' anonymity and confidentiality of transactions \cite{xu2017taxonomy}. In general, ZKP's aim to confirm a statement about a transaction such as "This is a valid transaction" without revealing anything about the transfer (statement) itself or the parties involved \cite{yu2018virtualization,xu2017taxonomy, globalblockchain2017}. 

Zerocoin was the first initiative with the aim of providing transaction unlinkability using ZKP's \cite{dinh2018untangling}. Similar to the Bitcoin Zerocoin uses the PoW consensus protocol to validate transactions. A \emph{cryptographic mixer} is implemented for Zerocoin to conceal the links between a zerocoin and the corresponding Bitcoin. Mixing services group multiple transactions in such a way that a payment contains several input addresses and several output addresses \cite{xu2017taxonomy}. When validating the transaction the miners do not have to verify the transaction with a digital signature, but rather validate whether coins belong to a list of valid coins \cite{mingxiao2017review}. Transaction unlinkability is achieved because the coin exchanged by the zerocoin owner can be any of the listed zerocoins \cite{dinh2018untangling, xu2017taxonomy}. A series of mixing service can be linked to one another to further improve transaction unlinkability. If the mixed transactions are equal in value this further minimizes the traceability between the input and output addresses \cite{xu2017taxonomy}. There are \emph{centralized mixing services} that require a third party to enact the mixing, and \emph{distributed mixing services} that do not require a third party to mix the coins. Coinshuffle \cite{ruffing2014coinshuffle} for instance, is a decentralized mixing protocol that is operated without any third party incurring only small additional overhead. Despite the fact that the cryptographic mixer unlinks the origins of the transactions to prevent transaction graph analysis the transactions' destination and amounts are still traceable \cite{mingxiao2017review}.

Building on the ZKP approach as a foundation, Zcash, extent the privacy guarantees, and improve the efficiency (throughput and latency) of Zerocoin. Zcash uses a variant of the PoW called Equihash. Transactions made using Zcash, including the split and merge transactions, are fully private \cite{dinh2018untangling}. Zcash employs a technique called \emph{Zero-Knowledge-Succinct Non-Interactive Argument of Knowledge} (zk-SNARKS) to provide these privacy guarantees \cite{yu2018virtualization, mingxiao2017review} that are a specific type of ZKP. The transactions are based on a complex concept of ZKP that reveal only that unredeemed coins exists the sum of which is a certain value other information with regard to the origin and destination address will remain concealed \cite{centralbankofbrazil2017distributed}. Blockchains that implement zk-SNARK techniques incur large overheads due to the fact that ZKP's require public parameters that when stored can amount up to hundreds of megabytes \cite{dinh2018untangling, internetofthingschristidis2016blockchains}.

\subsection{Physical View}
\label{app_physicalview}
The physical view is concerned with the topology of software components and their physical connections. Electronic devices known as nodes constitute to a blockchains' P2P network and are the only physical connection to the non-digital world. P2P networks on which blockchain platforms are run have different arrangements; First, the network can be categorized on the basis of permissions (authorization). Second, networks can be categorized with regard to their accessibility. Permissions to perform operations on the blockchain might differ ranging from allowing anyone to read, write and to partake in the consensus protocol to only one of these permissions. Control over these permissions can be confined to a distinct group of nodes, or all nodes. 

\subsubsection{Permissionless Blockchain Models}
\label{app_Permissionless Blockchain Models}

As the name suggest \textit{Permissionless} grant permission to all nodes in the P2P network to read and write transactions. By design permissionless blockchains are highly decentralized platforms, and permissions for all nodes are equal. These permissions were pre-defined when designing the blockchains architecture \cite{yaga2018blockchainoverview, DoyouneedaBlockchain?2017}. Granting write and read permission to all nodes that participate in the P2P network allows anyone to contribute data to the distributed ledger, and obtain an identical copy of the ledger. It assures \textit{censorship resistance}, preventing that an actor can block a valid transaction from being added to the ledger \cite{walport2016distributed, gatteschi2018blockchain, deblockchainsforgovernmental2017distributed}. Furthermore, since multiple nodes hold a copy of the ledger and are permitted to read the transaction data, these systems are highly \textit{transparent} \cite{Zheng2017AnOverviewofBlockchainTechnologyArchitectureConsensusandFuturetrends}. Permissionless blockchains are designed in this fashion to ensure that all participants in the network have a \textit{consistent view} of the distributed ledger \cite{globalblockchain2017}. The transparency of permissionless blockchains has a trade-off: It inherently means that individual nodes have less privacy with regards to their transactions. Currently most permissionless systems offer only \textit{pseudo anonymity}, which means that the individual users are anonymous but as their transactions are transparent their accounts are not \cite{BlockChainTechnologyBeyondBitcoin2015, centralbankofbrazil2017distributed, DistributedLedgerTechnologyandBlockchain2017, swan2015blockchain}. Indeed, the origins of a transaction and their destination can be revealed \cite{van2017process,biryukov2014deanonymisation}, and some research even lay bare behavior of Bitcoin users, how they spend their Bitcoins and the balance of Bitcoins they keep in their account.

However, permissionless blockchains ensures that transactions once recorded on their ledgers become nearly \textit{immutable} \cite{deblockchainsforgovernmental2017distributed}. Immutability means that transactions once recorded on the distributed ledger can not be erased or changed \cite{gatteschi2018blockchain, applicationsofdistributed2017} and only novel transactions can be appended \cite{yaga2018blockchainoverview}. A major drawback of permissionless systems is that in the event that security issues arise, or unauthorized transactions are executed, the response in the form of an update might be slow due to the hampered decision making \cite{deblockchainsforgovernmental2017distributed}. 

\subsubsection{Permissioned Blockchain Models}
\emph{Permissioned} blockchain platforms have confined and idiosyncratic permissions for their nodes \cite{xu2017taxonomy, yaga2018blockchainoverview, distributedledgertechnologyinpayments2016, arevolutionintrust2017}. As such permissioned blockchain platforms offer high utility for organizations that require control over important functions. However, privacy of transactions is at the expense of immutability guarantees because confining the ability of every node to write transactions creates censorship resistance, whereas limiting the read ability of nodes in the network compromises transparency \cite{deblockchainsforgovernmental2017distributed}. Since this model also requires a single entity that regulates permissions that could become a potential target of a cyberattack itself \cite{DistributedLedgerTechnologyandBlockchain2017}. Another advantage of permissioned models is that updating the blockchain can be delegated to a select group of nodes. This reduces the time it will take to implement the update \cite{cachin2017blockchains,xu2016connectblockchain}. In the event of any malfunctioning of the system the response time will therefore be significantly lower as compared to permissionless systems where multiple nodes need to come to an agreement. 

The P2P network can also be described from the perspective of network accessibility. In the literature three categories of P2P networks can be distinguished that are coupled to a permission model introduced in the previous paragraph \cite{Zheng2017AnOverviewofBlockchainTechnologyArchitectureConsensusandFuturetrends,xu2017taxonomy,globalblockchain2017,deblockchainsforgovernmental2017distributed,de2017understanding,cachin2017blockchains}. 

\subsubsection{Public Blockchain Networks}
A \emph{Public blockchain}, like the Bitcoin or Ethereum have \emph{open network access} meaning that anyone willing is allowed to join the network. Public blockchains have no single owner. In fact they can not be owned by one single entity \cite{distributedledgertechnologyforthefinancialindustry2016, yaga2018blockchainoverview}. Therefore, these networks can be regarded as \emph{decentralized} \cite{Zheng2017AnOverviewofBlockchainTechnologyArchitectureConsensusandFuturetrends}. Since the nodes do not know one another, they cannot trust each other \cite{swan2015blockchain, cachin2017blockchains, xu2017taxonomy}. This assumed \emph{trustless} setting has important ramifications for the design of a public blockchains architecture \cite{deblockchainsforgovernmental2017distributed}. For instance, Bitcoin was designed to assume that any node participating in the network needs to be distrusted, but the system will be secure for as long as the majority of nodes act honestly during the consensus protocol \cite{embracingdisruption2016}. The Bitcoin and other public blockchains have to employ a complex and computational expensive consensus protocol to incentivize and regulate cooperative behavior between nodes \cite{deblockchainsforgovernmental2017distributed, dinh2017blockbench, cachin2017blockchains}. Permissionless models are the hallmark of public blockchains \cite{arevolutionintrust2017}. The combination of open network access and a permissionless systems permits nodes that are part of the P2P network to perform all operations available. Consequently, for public blockchains transparency of transactions on the distributed ledger is paramount, and is as such embedded in the blockchains architecture \cite{globalblockchain2017}.

\subsubsection{Private Blockchain Networks}
\emph{Private blockchains} are blockchains networks that are owned by one organization. Blockchain networks that have one or few owners, and for which the permissions of the blockchain are vetted by a confined group of nodes can be regarded as more \emph{centralized} \cite{Zheng2017AnOverviewofBlockchainTechnologyArchitectureConsensusandFuturetrends}. These type of blockchains have limited network access which allows access only for authenticated nodes \cite{dinh2017blockbench, arevolutionintrust2017, deblockchainsforgovernmental2017distributed}. One gatekeeper, or a selection of nodes might have the permission to grant access to the network \cite{globalblockchain2017}. Another option is that only predefined nodes that are white listed can enter the network \cite{DoyouneedaBlockchain?2017,dinh2018untangling,internetofthingschristidis2016blockchains}. In essence, both options entail that before entering the network nodes are \emph{authenticated} \cite{dinh2017blockbench, arevolutionintrust2017}. 

Private blockchain platforms require that an administrator, or administrators assign permissions to each unique node \cite{yaga2018blockchainoverview}. The chances of nodes acting malicious are mitigated as their identity is known by the other nodes \cite{deblockchainsforgovernmental2017distributed}. This does not imply however, that the nodes in the private blockchain network can fully trust each other \cite{dinh2018untangling}. Because the identities of the nodes are known, the risks of Sybil attacks are slim. Computational intensive (thus expensive) consensus algorithms such as PoW or PoS are therefore less needed for the system to become byzantine fault tolerant \cite{dinh2017blockbench,internetofthingschristidis2016blockchains,DistributedLedgerTechnologyandBlockchain2017}. 

\subsubsection{Consortium Blockchain Networks} \emph{Consortium blockchains} are similar to private blockchains in the sense that nodes first need to be authenticated before granted access to the network. However, consortium blockchains allow nodes from different organizations to access the blockchain network \cite{xu2017taxonomy, mingxiao2017review,pilkington2016}. In addition, a consortium blockchain is usually owned by more than one entity. As such the degree of openness and centralization of consortium blockchains lies between a private and public blockchain \cite{mingxiao2017review}. Within consortium blockchain P2P networks the consensus process is confined to specific participants \cite{xu2017taxonomy, peters20166understanding, Zheng2017AnOverviewofBlockchainTechnologyArchitectureConsensusandFuturetrends}. The right to read transactions stored on the blockchain network can be made public within the network, or restricted to a selected group of nodes \cite{deblockchainsforgovernmental2017distributed, xu2017taxonomy}. Consortium blockchains can therefore be perceived as being \emph{partially decentralized}. Following similar reasoning as for private blockchain, the need for computationally expensive consensus algorithms is less present.

%% file: appendix_applications.tex
\subsection{Cryptocurrencies}
\label{app_cryptocurrency}

Cryptocurrencies can be regarded as the vanguard of BCT and were its first application \cite{swan2015blockchain}. The inhibition of updates to introduce novel features and various other reasons sparked the creation of other cryptocurrencies besides the Bitcoin known as \emph{altcoins}. These altcoins share a common code base with the Bitcoin \emph{parent chain} except for the code enabling the novel functions these coins aim to offer \cite{back2014enabling, zohar2015bitcoin}. To implement the additional codebase these altcoins branch off in an alternative chain or \emph{altchain}, from the Bitcoin's main chain effectively forming a new blockchain. Since these new blockchains (altchains) share at least a common code base for the implementation of proof of work consensus protocol they can share miners with their Bitcoin parent chain to mine blocks - this process is called \emph{merged mining} \cite{BlockChainTechnologyBeyondBitcoin2015}. When engaging in merged mining, miners concurrently try to find a PoW solutions for multiple cryptocurrencies. The concept of merged mining was introduced as a solution to diminish the possibility of fragmentation the mining power among competing cryptocurrencies and to bootstrap new cryptocurrencies \cite{voyiatzis2017merged}.

The emergence of altcoins enabled the introduction of novel blockchain functions yet it also raised two problems \cite{back2014enabling}. First, although merged mining mitigates mining power fragmentation, altcoins did lead to infrastructure fragmentation since each uses its own technology stack in addition to their shared codebase. The codebase of the altchains often inhibits the implementation of security updates. Therefore efforts to improve the underlying technology are regularly unnecessary duplicated or are lost. Second, cryptocurrencies typically have a floating price (expressed in fiat currency) \cite{mai2018does, li2017technology} that might differ from one another. When opting to exchange altcoins for one another, or a Bitcoin the owners of a coin have to resort to a centralized trusted third party such as \emph{decentralized exchanges} (DEX). A DEX also facilitate the exchange of fiat currencies (e.g. Dollar or Euro) for cryptocurrency such as Ether (native cryptocurrency of Ethereum) or Bitcoins. In their work Decker et al. \cite{decker2015making} propose an automated software-based audit that aids users of a DEX in determining the solvency of the DEX without revealing any private data.

Despite the efforts to improve the trustworthiness of decentralized exchanges both the aforementioned problems persisted to hamper the \emph{interoperability} between blockchains and thus further innovation and creation of the blockchain ecosystem. The concept of \emph{pegged sidechains} to automatize and secure the transfer of assets between one chain and another \cite{back2014enabling}. A pegged side chaining mechanism is two-way; it allows for the transfer of tokens back and forth to the original chain. The original chain is called the parent chain, and the one to which the coins are transferred to as the \emph{sidechain}. To secure the transfer the parent chain coins are sent to a special output on the parent chain that can only be unlocked by a proof of possession on the sidechain \cite{back2014enabling}. Sidechains are backed by Bitcoins via a Bitcoin contract similar like how dollars and pounds used to be backed by gold \cite{BlockChainTechnologyBeyondBitcoin2015}. 

\subsection{Smart Contracts}
\label{app_smartcontracts}
The term smart contract was first pioneered by Nick Szabo \cite{szabo1997formalizing} in 1997. Central to the idea of smart contracts is that every computable clause can be encoded into computer programs, and let the program decide and execute what happens during the contract's life span \cite{frowis2017code}. Business logic and conditions can be expressed that allow for complex transactions which are executable on a blockchain \cite{xu2017taxonomy}. Ethereum is the most widely used platform to deploy smart contracts \cite{xu2017taxonomy}. To support user in writing their smart contracts Ethereum supports several high-level programming languages that are \emph{turing complete} such as Serpent, LLL, and Solidity \cite{arevolutionintrust2017, luu2016making, EthereumHomesteadDocumentation2017}. These high level language smart contracts are then compiled into bytecode that can be executed in an environment called the Ethereum Virtual Machine (EVM), which can be regarded as the core of the Ethereum platform \cite{EthereumHomesteadDocumentation2017}.

On the Ethereum platform transactions can invoke smart contracts. Transactions made to smart contracts contains a \emph{startgas} value that express the maximum number of computational steps the transaction is allowed to execute, and the coherent \emph{gasprice value} that represents the fee a sender is willing to commit in order these steps. Each gas unit committed allows for the the execution of an atomic instruction that reflects a computational step \cite{luu2015demystifying, EthereumHomesteadDocumentation2017}. Gas was introduced as a transaction fee for the execution of smart contracts to make distributed denial of service attacks expensive \cite{deblockchainsforgovernmental2017distributed}. 
Besides the possibility of denial of service attacks, research by Luu et al. \cite{luu2016making} suggests that Ethereum smart contracts are susceptible to abuse of miners and contract users. In practice they suggest, there is a \emph{semantic gap} between the assumptions contract writers make about the semantics of a smart contract and the actual semantics of the contract. In their study they suggest several improvements, and a tool to check for vulnerabilities in Ethereum smart contracts. Programs stored on the Ethereum blockchain are considered immutable since they are recorded on a blockchain. However, it has been shown that the control flow of these smart contracts can be changed to exploit semantic loopholes \cite{frowis2017code}. The execution and verification of smart contracts is performed sequentially which hampers their throughput and latency. Dickerson et al. \cite{dickerson2017adding} propose a novel way that permits miners and validators to execute smart contracts in parallel so that the process of execution and verification can be performed faster.  

Another pervasive problem for smart contracts is that the outputs resulting from their execution is transparent and visible to the network. In \cite{kosba2016hawk}, Hawk a blockchain model to preserve the privacy of smart contracts is introduced. The model extends the Zcash protocol (see \ref{zkp}) by adding to the possibility to stipulate arbitrary transaction logic to write smart contracts. In addition, for Zcash local states do not need to be protected whereas this is required when executing smart contracts. Hawk processes the transaction inputs and outputs to conceal any cryptographic details. 

\subsection{General Applications}
\label{app_generalapp}

Our data suggests that general-purpose applications of BCT can be arranged into five additional categories.

\subsubsection{Security and Fraud Detection}

For instance, Cruz et al. \cite{cruz2018rbac} present a platform that uses Ethereum smart contracts to realize transorganizational Role Based Access Control. BCT could be applicable to various use cases in the field of intrusion detection \cite{meng2018intrusion}. BCT could resolve the vulnerabilities that compromise mobile applications of cloud-based messaging solutions often provided by cloud service providers, and mobile systems manufacturing firms \cite{esposito2018cloud}. The security of robotic swarm systems when facing byzantine robots could also be managed using BCT \cite{strobel2018managing}.  

The inherent transparency and tamper proof transactions that BCT can serve for \emph{fraud detection}. To prevent fraud in the shape of event (e.g. concerts) ticket theft, selling multiple copies of one ticket, or selling invalid tickets \cite{tackmann2017secure}. Another interesting application for BCT related to fraud detection is a Linked Data based approach to create audit logs that provide proof of log manipulation and non-repudiation by privacy auditors. This approach has the aim to protect the privacy of individuals who contribute their data to a collaborative service or research environment \cite{sutton2017blockchain}. In the work \cite{zhao2018secure} a new system is proposed to improve traditional pub-sub (publish-subscribe) models for sharing data that rely on publisher and subscriber confidentiality and anonymity. The secure pub-sub model (SPS) opts to eliminate the need for middle-ware with BCT based fair payment mechanisms based on data source reputation on a blockchain. 

\subsubsection{Securities and Insurance} 

Several practitioners view how BCT could be utilized for \emph{financial securities} \cite{distributedledgertechnologyforthefinancialindustry2016,distributedledgertechnologyinpayments2016,DistributedLedgerTechnologyCybersecurity2016} this might not be surprising since BCT was originally invented for financial transactions. Recently the European Central Bank and the Bank of Japan have engaged in a joint research project, STELLA, to explore how the delivery process of securities against the payment of cash could be designed and operated using distributed ledger technology \cite{Securitiessettlementsystems2018}. In one of their papers \cite{distributedledgertechnologyinpayments2016} the Federal Reserve expounds how DLT or BCT might be used for payments, clearing and settlement. Challenges and opportunities that these technologies face for their long-term adoption are also discussed. The Central Bank of Brazil \cite{centralbankofbrazil2017distributed} has conducted an experiment to use BCT for an interbank payment schema. The aim of the expermiment was not to substitute or replace current settlement systems, but rather to investigate whether it could be a back up system in case the normal payment system would fail. In \cite{vo2017blockchain} an insurance application is presented that utilizes BCT and smart contracts for \emph{insurance} contracts. More specifically, to offer a transparent insurance quote to customers of insurance firms. 

\subsubsection{Record-Keeping}

The blocks that constitute to a chain encompass a record of transactions that can be utilized to track various types of the assets such as intellectual property, digital identities and electricity. These applications are not necessary focused on the transaction itself, but rather emphasize on recording and historically preserving records. Ekblaw et al. \cite{Ekblaw2017Acase} for instance introduce a prototype to electronically store \emph{health records} and research data. Azouvi et al.  \cite{azouvi2017secure}, Lin et al. \cite{lin2018new} and Lee \cite{lee2018bidaas} introduce a blockchain based solution to record \emph{digital identities} bind them to their real world entity, registering identities and their attributes, and authentication management. For \emph{energy records} BCT could be utilized to record a consumers' energy consumption in real time \cite{pop2018blockchain}. Can provide consumers with more insight into the energy consumption of their appliances. A prototype platform to register consumer energy consumption while simultaneously mitigating the risk that energy consumption data of consumers falls into the wrong hands has been developed \cite{gao2018gridmonitoring}. Huang et al. \cite{huang2018lnsc} present a decentralized security model that is based on the lightning network \cite{bitcoinlightning} in combination with smart contracts to secure the management between electronic vehicles and charging piles. 

Another application for BCT is \emph{supply chain management}. For instance it could be used to track various goods to determine whether they are counterfit and their origins \cite{toyoda2017novel}. Korpela et al. \cite{korpela2017digital} investigate the requirements of supply chain integration and provide a broad insight how BCT can transform digital supply chains and networks. Besides these applications Turkanovi{\'c} \cite{turkanovic2018eductx} introduce a global higher education credit platform (EduCTX) to offer a globally unified viewpoint for students and higher education institutions. The platform to process, manage and control ECTX tokens that represent credits that students will obtain for completing courses similar to ECTS that are currently being used. BCT can also be employed for other academic purposes such as peer-reviewing. To this end Hoffman et al. \cite{hoffman2018smart} introduce a blockchain based smart contracts to track the social interactions for scholarly publications in a decentralized manner. 

\subsubsection{Internet-of-Things}

Devices connected to the \textit{Internet of Things} (IoT) can generate vast amounts of data that could potentially provide invaluable information for the execution of smart contracts. According to Cha et al. \cite{cha2018blockchain} an dedicated Internet of Things (IoT) smart contract could be used as a gateway service to manage the IoT device's information and privacy policies. By using the gateway users can access their IoT device to obtain information and privacy policies. In \cite{internetofthingschristidis2016blockchains} it is argued that smart contract can offer support for the IoT by cutting the cost of updating an IoT device. For instance by offering the firmware at first via a node owned by the manufacturer that is part of the network. Daza et al. \cite{daza2017connect} propose a discovery approach to let IoT devices become aware about their surrounding environment. By means of a discovery scheme that helps identifying IoT services and, when required, the device that runs this service. In \cite{sharma2018secure} a novel DMM (Distributed Mobility Management) schema is proposed based on blockchain technology with the aim of resolving hierarchical security issues without affecting the IoT network layout. The work by Patil et al. \cite{patil2017framework} presents a security framework and lightweight BCT architecture which allows IoT devices in greenhouse farms to act as a centrally managed blockchain with decentralized security and privacy. The security framework presented provides a secure communication platform for IoT devices. 

\subsubsection{Smart Legal Contracts}
Although the term smart contracts includes the word "contract" that does not mean that they are always legally enforceable in court. In their works Clack et al. \cite{clack2016smart,clack2016smartfoundations} set out a vision how smart contracts can become legally enforceable, thus creating \emph{smart legal contracts}. A precursor to such contracts is presented by Pinna et al. \cite{pinna2017blockchain} that propose a BCT-based system with the aim of enacting temporary employment contracts that protect employees against insolvency of the employer.